\documentclass[manuscript]{aastex}

\usepackage{bm}
\usepackage{amsmath}
\usepackage{mydef}

\slugcomment{For submission to PASP}

\shorttitle{Submillimeter polarimetry with PolKa}
\shortauthors{Wiesemeyer et al.}

\begin{document}


\title{Submillimeter Polarimetry with PolKa, a reflection-type modulator for the APEX telescope} 


\author{H. Wiesemeyer\altaffilmark{1},
        T. Hezareh\altaffilmark{1}, 
        E. Kreysa\altaffilmark{1}, 
        A. Weiss\altaffilmark{1},
        R. G{\"u}sten\altaffilmark{1},
        K.M. Menten\altaffilmark{1},\\
        G. Siringo\altaffilmark{2},
        F. Schuller\altaffilmark{2},
        A. Kovacs\altaffilmark{3, 4}}

\affil{(1) Max-Planck-Institute for Radio Astronomy, Bonn, Germany\\
       (2) European Southern Observatory, Santiago, Chile \\
       (3) University of Minnesota, Minneapolis, MN 55405 \\
       (4) California Institute of Technology, Pasadena, CA 91125
}
\email{hwiese@mpifr.de}




\begin{abstract}
Imaging polarimetry is an important tool for the study of cosmic magnetic
fields. In our Galaxy, polarization levels of a few up to $\sim$10\%
are measured in the submillimeter dust emission from molecular clouds and in the
synchrotron emission from supernova remnants. Only few techniques exist
to image the distribution of polarization angles, as a means of tracing the
plane-of-sky projection of the magnetic field orientation. At submillimeter wavelengths,
polarization is either measured as the differential total power of
polarization-sensitive bolometer elements, or by modulating the polarization of
the signal. Bolometer arrays such as LABOCA at the APEX telescope are used to
observe the continuum emission from fields as large as $\sim0\fdg2$ in diameter.
Here we present PolKa, a polarimeter for LABOCA with a reflection-type waveplate
of at least 90\% efficiency.
The modulation efficiency depends mainly on the sampling and on the angular
velocity of the waveplate. For the data analysis the concept of generalized
synchronous demodulation is introduced. The instrumental polarization towards
a point source is at the level of $\sim0.1$\%,
increasing to a few percent at the $-10$db contour of the main beam.
A method to correct for its effect in observations of extended sources
is presented. Our map of the polarized synchrotron emission from the Crab
nebula is in agreement with structures observed at radio and optical
wavelengths. The linear polarization measured in OMC1 agrees with
results from previous studies, while the high sensitivity of LABOCA
enables us to also map the polarized emission of the Orion Bar, a prototypical
photon-dominated region.
\end{abstract}





\section{Introduction}
\label{sec:intro}
Magnetic fields are important constituents in the interstellar medium
(ISM) and are theoretically shown to control many physical processes
including formation and fragmentation of molecular clouds (e.g.,
\citealp{2005ApJ...633L.113H, 2008A&A...486L..43H}), and regulation of the
process of star formation (e.g., \citealp{1999osps.conf..305M}).
Observational studies, on the other hand, are essential to
provide constraints on the existing numerical models. However,
magnetic field observations require state-of-the art instruments and
intricate observing and data reduction techniques.

Submillimeter polarimetry is one of the best tools for characterizing 
interstellar magnetic fields, because in most environments commonly
observed aspherical dust grains are aligned with their principle
axis of major inertia along the magnetic field lines. The physics
of grain alignment is extremely rich and complex; \citet{2007JQSRT.106..225L}
and (with emphasis on cold clouds) \citet{2004tcu..conf..213D} review the underlying
theory. For decades dissipation of rotational energy in paramagnetic grains \citep{1951ApJ...114..206D},
set into suprathermal rotation by various torques \citep{1979ApJ...231..404P}, was considered the dominant
process. Towards the end of the 90s it became clear that anisotropic
radiative torques do not only spin up the grains \citep{1996ApJ...470..551D}, but also align them
efficiently with respect to the magnetic field \citep{1997ApJ...480..633D}. This alignment mechanism works
for larger grains with effective radii $a_{\rm eff} \ga 0.1\mu$m, while smaller grains are at most weakly aligned
\citep{1995ApJ...444..293K,2009ApJ...696....1D}. More recent work on radiative torques is
reviewed by \citet{2011ASPC..449..116L} with further references therein. We also note that dust grains
exposed to sub- or super-sonic gas flows are expected to be mechanically aligned with their long axis
perpendicularly to the magnetic field \citep{2007ApJ...669L..77L}. In summary, for molecular clouds the 
thermal emission of dust is linearly polarized orthogonally to the field lines,
and a polarization map will reveal the direction of the plane-of-sky component of the
magnetic field. For equipartition between magnetic pressure and turbulent and thermal
presure, its strength can be determined by means of the Chandrasekhar-Fermi method \citep{1953ApJ...118..113C}.
This technique was originally applied to estimate the magnetic field strength
in Galactic spiral arms. Chandrasekhar and Fermi used the dispersion of
polarization vectors with respect to the large-scale magnetic field, known, from
the dichroic absorption of starlight \citep{1951ApJ...114..241H}, to be
generally aligned with the Galactic plane.
\citet{2009ApJ...696..567H} and \citet{2009ApJ...706.1504H} have developed the
CF-method further, accounting for the distortion of the magnetic field by turbulence; their
analysis does not need to assume any model for the large-scale field. MHD simulations
have shown (\citealt{2001ApJ...561..800H}) that modified CF-methods are
astonishingly robust, and that order-of-magnitude estimates are possible even in
regions that are not dominated by hydromagnetic turbulence. 

This work deals exclusively with linear polarization. While in the NIR a substantial
fractional circular polarization can be expected for light scattered by aligned dust grains
($\simeq 15$\% have been observed in the BN object in OMC1,
\citealt{2009ApJ...697..807M}) or in the \wl{3.1} feature of water ice (e.g.,
\citealt{2006MNRAS.366..491A} for a summary of detections and further references),
at sub-millimeter wavelengths we expect at most a weak signal. --
In practice, the polarimetry of dust emission is completed by
analyzing the circular and linear polarization of suitable spectral lines
(Zeeman and Goldreich-Kylafis effects, for a review see e.g.,
\citealp{2012ARA&A..50...29C}). 

Sub-millimeter imaging polarimetry is not limited to dust polarization, but
also useful to study the synchrotron emission from e.g., plerion-type supernova
remnants. While polarimetry reveals the magnetic field structure, observations
into the far-infrared allow to determine
the break in the spectral energy distribution and therefore the strength of the 
magnetic field \citep{1984ApJ...278L..29M,1989ApJ...338..171S}, and to trace
the production of dust by the progenitor star \citep{2004MNRAS.355.1315G}.

As the theory and analytical methods evolved, technological progress has been
made in submillimeter imaging and polarimetry. Since low-mass protostars  
in nearby ($d\sim100$~pc) star forming regions and protoclusters in regions
where massive stars form ($d\sim 1$~kpc) extend over scales corresponding to
the limited field-of-view of (sub)millimeter
interferometers (at most a few minutes of arc), remarkable results have been
achieved thanks to their dual-polarization
capabilities and powerful digital cross-correlators (e.g.,
\citealt{2013ApJ...772...69G}). On larger scales, inaccessible to
interferometers, the imaging surveys done with the Herschel space observatory
(e.g., a survey of the Gould Belt in which low-mass stars
form, \citealt{2010A&A...518L.102A}, and the HOBYS survey for regions forming
massive stars, \citealt{2010A&A...518L..77M}), revealed filamentary structures
across large areas on the sky. While the formation of these filaments out of
the warm, neutral phase of the interstellar medium has motivated the
aforementioned and other theoretical studies (e.g., \citealp{2005ApJ...633L.113H, 2008A&A...486L..43H}),
the large-scale magnetic field and therefore its dynamical role are still poorly constrained by observations. In the past
decades, several wide-field bolometer cameras at various submillimeter observatories
have therefore been equipped with polarimeters. Table~\ref{tab:polarimeters}
summarizes their main features along with published first-light
observations. Submillimeter interferometers are not considered because they are 
blind to the large spatial scales addressed by this paper. Likewise,
spectro-polarimeters are omitted because they are designed for different, yet often
complementary studies (e.g., correlation polarimeters such as XPOL, \citealp{2008PASP..120..777T}).
\begin{table*}[ht!]
\begin{center}
\caption{Submillimeter imaging polarimeters (past and present)$^{(a)}$.}
\label{tab:polarimeters}
\begin{tabular}{lllcccl}
\tableline\tableline
telescope & camera & polarimeter & field of view$^{(b)}$
& wavelength$^{(c)}$ & beam$^{(d)}$ & method \\
          &           &             &  [\arcmin] & [$\mu$m] & [\arcsec]  & \\
\tableline
KAO  & \multicolumn{2}{c}{\Vhrulefill\,\,Stokes$^{(1)}$\,\,\Vhrulefill} & 3.1 & 100 & 35 & quartz \\
CSO  & \multicolumn{2}{c}{\Vhrulefill\,\,Hertz$^{(2,3)}$\,\,\Vhrulefill} & 1.8 & 360, 450 & 20 & quartz \\
JCMT & SCUBA$^{(4)}$ &  SCUPOL$^{(5)}$ & 2.3 & 450, 850 & 8 & quartz \\
CSO  & SHARC-II$^{(6)}$ & SHARP$^{(7)}$ & 2.4 & 350, 450 & 20 & quartz \\
SMTO & Hertz$^{(2)}$ & VPM$^{(8)}$ & 1.8 & 350 & 20 & dual VPM\\
Planck & \multicolumn{2}{c}{\Vhrulefill\,\,HFI$^{(9,10)}$\,\,\Vhrulefill} & n.a. & 3000, 2100, 1380, 850 & 282 & PSB\\
\multicolumn{3}{c}{\Vhrulefill\,\,BLAST-Pol$^{(11)}$\,\,\Vhrulefill}  & $13.5 \times 6.5$ & 250, 350, 500 & 30  & PSB \\
APEX & LABOCA$^{(12)}$ & PolKa$^{(13)}$ & 12.3 & 870 & 20 & RPM \\
\tableline
\end{tabular}
\tablecomments{(a) Cameras are specified separately when subsequently equipped
with a polarimeter. Only commissioned, published systems are considered.
(b) Largest dimension of instantaneous field-of-view (refers to the shortest
wavelength for detectors with several bands). (c) Central wavelength of camera
bandpass. (d) {\sc fwhm} of main beam in Stokes $I$, not necessarily the diffraction
limit of the telescope. Refers to shortest wavelength for polarimeters with several bands.
{\bf References:} 
(1) \citet{1991PASP..103.1193P},
(2) \citet{1997PASP..109..307S},
(3) \citet{1998ApJ...504..588D},
(4) \citet{1999MNRAS.303..659H},
(5) \citet{2003MNRAS.340..353G},
(6) \citet{2003SPIE.4855...73D},
(7) \citet{2008ApOpt..47..422L},
(8) \citet{2008ApOpt..47.4429K},
(9) \citet{2010A&A...520A...9L},
(10) \citet{2010A&A...520A..13R},
(11) \citet{2010SPIE.7741E...9F},
(12) \citet{2009A&A...497..945S},
(13) \citet{2012SPIE.8452E..06S}.\\
{\bf Acronyms:} 
APEX - Atacama Pathfinder Experiment (Llano Chajnantor, Chile),
BLAST-Pol - Balloon-borne Large Aperture Submillimeter Telescope for
Polarimetry, 
CSO - Caltech Submillimeter Observatory (Mauna Kea, Hawaii),
HFI - High Frequency Instrument,
JCMT - James Clerk Maxwell Telescope (Mauna Kea, Hawaii),
KAO - Kuiper Airborne Observatory,
LABOCA - Large Apex Bolometer Camera,
PolKa - Polarimeter f{\"u}r Bolometerkameras,
PSB - polarization sensitive bolometer elements,
RPM - reflecting polarization modulator,
SCUBA - Submillimeter Common-User Bolometer Array,
SHARC - Submillimeter High Angular Resolution Camera,
SHARP - Submillimeter High Angular Resolution Polarimeter,
SMTO - Submillimeter Telescope Observatory (Mount Graham, Arizona),
VPM - variable delay polarization modulator.
}
\end{center}
\end{table*}

For incoherent detectors, such as in bolometer arrays, two methods to separate
polarized from unpolarized radiation are possible, namely (1) differential total
power measurements, and (2) modulations of the plane of polarization. In the
first technique, the total power of detectors that are sensitive
to polarizations at position angles $0\dg$, $\pm 45\dg$ and $90\dg$ yields
Stokes parameters $I$, $Q$ and $U$. This method is used in the low- and
high frequency instruments aboard the Planck satellite
(\citealp{2010A&A...520A...8L,2010A&A...520A..13R}), employing
polarization-sensitive elements \citep{2008SPIE.7020E..38K}, and for the
balloon-borne BLAST-Pol experiment \citep{2010SPIE.7741E...9F}.
In the second method the plane of polarization is rotated
periodically or stepwise, allowing to measure the polarization of the incident radiation field with
polarization-sensitive detectors, or, by means of a polarization analyzer (e.g., a
grid), also with detectors
otherwise insensitive to polarization. The modulator can be either a classical waveplate
operating in transmission or a tunable, reflecting polarizer (see Table~\ref{tab:polarimeters}).

In both methods the measurements of the Stokes parameters received from a given position
in the sky are not strictly simultaneous. A too slow detection cycle is inadequate especially
for ground-based observatories, plagued by atmospheric instabilities. In the first method,
the resulting 1/f noise can be suppressed by scanning fast enough. In the second method, the
rotation of the plane of polarization needs to be faster than the atmospheric total power 
fluctuation, suggesting a continuous rather than stepwise rotation. The suppression of
the 1/f noise in Stokes I, measured simultaneously with Stokes Q and U, still requires fast scanning.
We will come back to this topic in section~\ref{sec:reduction}.

This work describes PolKa \citep{2012SPIE.8452E..06S,2004A&A...422..751S},
the polarimeter for the Atacama Pathfinder Experiment (APEX, \citealp{2006A&A...454L..13G}). PolKa,
as a fast reflection-type polarization modulator, belongs to the second type of polarimeters. It is used 
with LABOCA, the Large APEX Bolometer Camera \citep{2009A&A...497..945S}. LABOCA operates at \wl{870} and
consists of 295 semiconductor bolometers, detecting the temperature rise due to the absorption of radiation.
The bolometers are arranged in an hexagonal grid with two beam spacing, providing an instantaneous
field-of-view of $0\fdg 2$. For the PolKa observations, 230 pixels offer an acceptable noise level.

The plan of the paper is as follows: Section \ref{sec:hardware} describes the hardware of the 
rotating half-waveplate, its working principle, the measurement
equation and the characterization and calibration of the device.  
Section \ref{sec:reduction} provides a detailed description of the data reduction methods such as
correlated-noise and bandpass filtering. We also introduce novel techniques
to demodulate the signals by a generalized synchronous demodulation and to remove
instrumental artifacts. These techniques are indispensable for imaging polarimetry and were
specifically created for PolKa. Section~\ref{sec:results} presents the first observational results
from PolKa with a reliable polarization angle calibration. We conclude the paper with
section~\ref{sec:conclusions} and refer the reader interested in the relevant mathematical methods to the appendix.
\section{Design of reflection-type polarization modulators}
\label{sec:hardware}
Polarization transforming reflectors were devised already more than two
decades ago \citep{1986IJIMW...7..1591H, 1988IJIMW...9..477P}, but in (sub-)millimeter astronomy
they have been used only twice. Several designations for devices of this kind exist in literature, for the
purpose of this paper we refer to them as reflecting polarization modulator (hereafter RPM).
\citet{1999PASJ...51..175S} and
\citet{2004A&A...422..751S} describe, respectively, the systems installed at the NRO 45m
(for spectral line polarimetry) and the SMTO 10m (with a 19-channel bolometer
array operating at \wl{870}) telescopes. Modulators of this kind have the double advantage
that their reflectivity is excellent (closer to 100\% than the transmission of 
classical waveplates) and that they can be tuned within a
wide wavelength range (only limited by the design of the grids).
\subsection{Working principle}
\begin{figure}
\begin{center}
\includegraphics[scale=0.56]{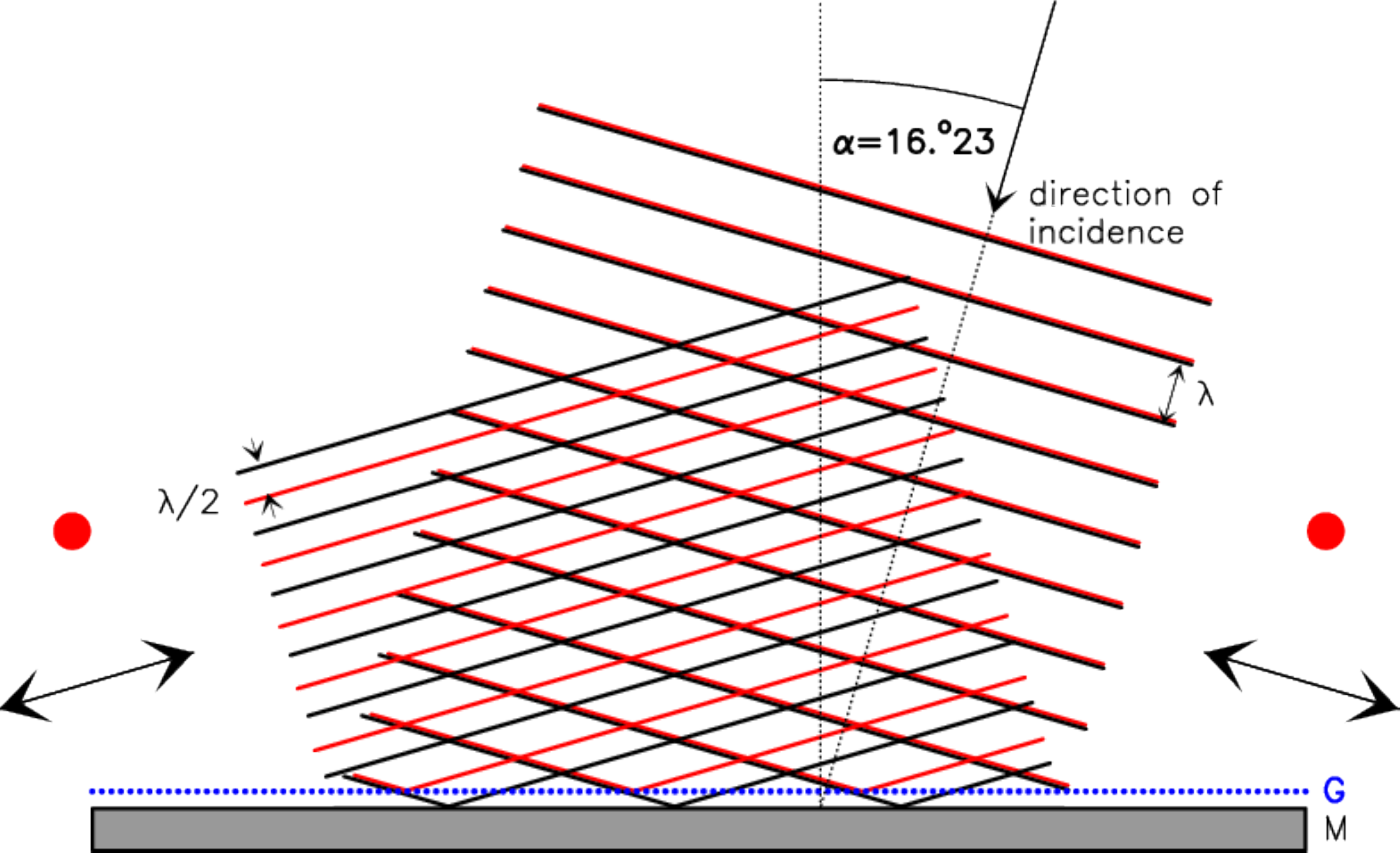}
\caption{Demonstration of wave propagation in a reflection-type polarimeter
with $\lambda$/2 tuning. The wave component with polarization in the drawing
plane is shown as black lines (representing the maxima of the wave front,
transmitted by grid G), that with polarization perpendicular to the drawing
plane as red lines (reflected by grid G). A pictorial representation of these
polarizations is also shown, for the incident wave (right) and the reflected
one (left). The grid (G) is shown by blue dots (wires perpendicular to the
drawing plane), and, at the bottom, the mirror (M) in gray. Not to scale. The
angle of incidence is given as for the installation of PolKa in the Cassegrain
cabin of the APEX telescope.}
\label{fig:scheme}
\end{center}
\end{figure}

RPMs are simple yet efficient optical equivalents to birefringent materials
(Fig.~\ref{fig:scheme}): In a parallel assembly of a grid and a mirror, mounted 
at an adjustable distance, the component of the incident radiation that is
polarized parallel to the wires can excite an electromagnetic mode and is reflected
off, while the component polarized perpendicular to the wires is
transmitted and then reflected into the outgoing ray where it is superposed to
the other ray, with a delay-induced phase difference. In a birefringent
material, these two components correspond to the ``fast'' and the ``slow'' rays,
respectively. If the device is tuned such that the phase difference between the 
two rays is $180\dg$, the overall effect is that of a classical half-waveplate,
i.e., transmitting both polarized and unpolarized radiation, but rotating the plane
of polarization by the double position angle of the grid. When a linearly
polarized incident signal is measured with a polarization-sensitive bolometer
element (or through an analyzer grid, for elements insensitive to polarization),
a rotation of the waveplate therefore modulates the received power. The amplitude
and the phase of the modulation (which for a constant rotation speed is sinusoidal)
then yields the linear polarization and its polarization angle, respectively.
%
%
\subsection{Description of the hardware}
\begin{figure*}[ht!]
\includegraphics[scale=0.8]{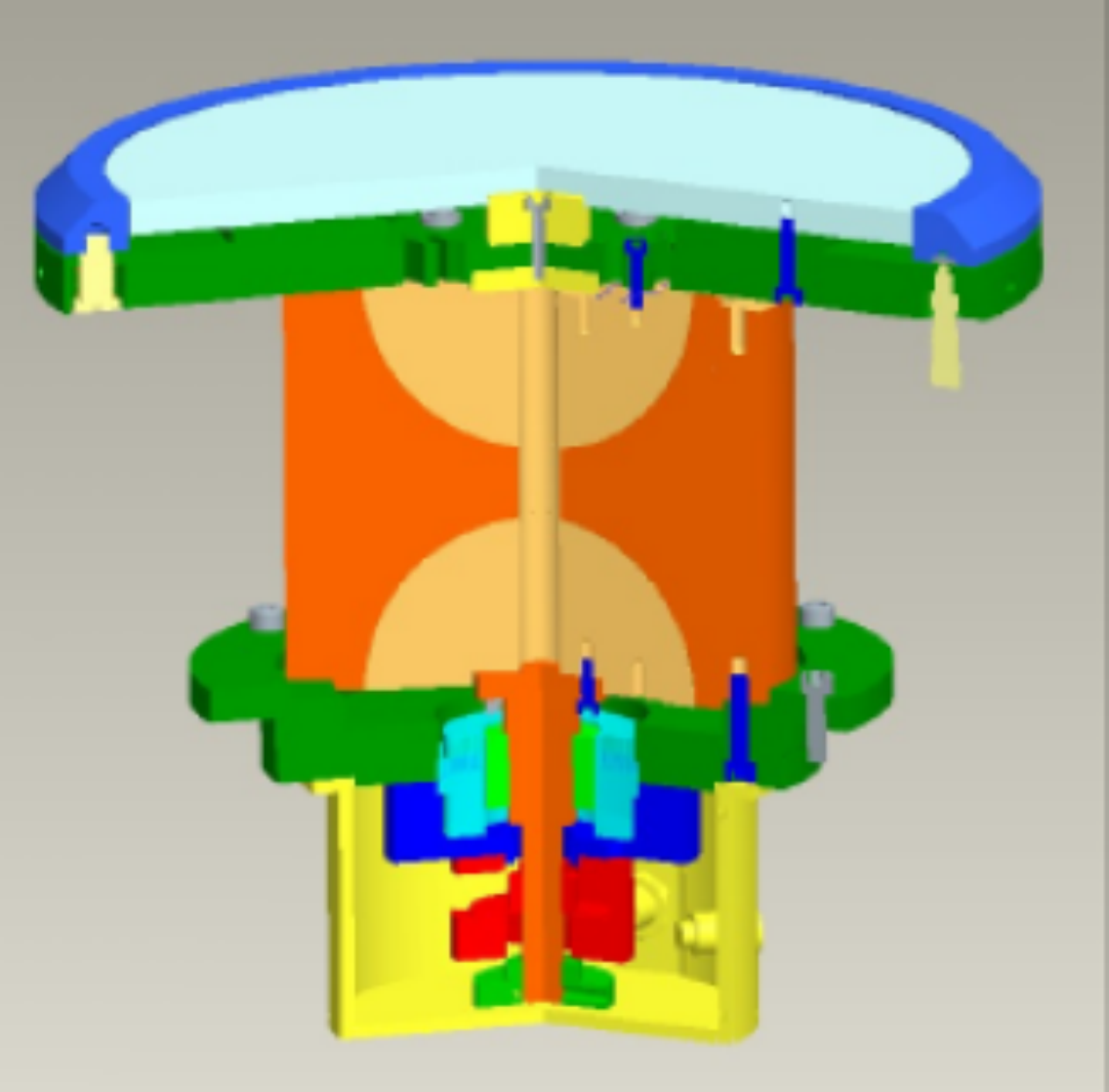} 
\hspace{.3cm}
\includegraphics[scale=0.621]{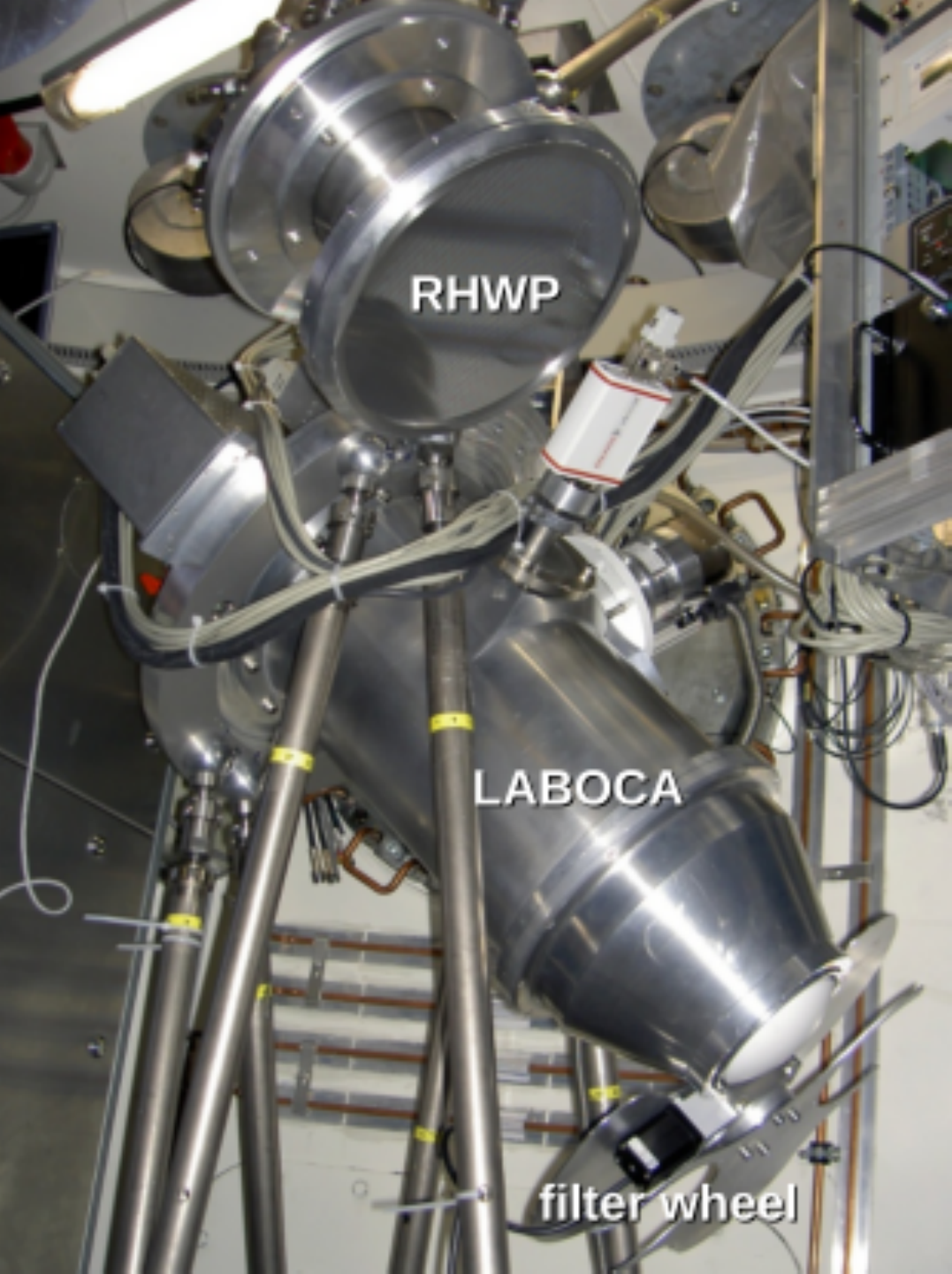} 
\caption{{\bf Left:} Design drawing of PolKa (reproduced with permission from
Dr. Stefan Risse; copyright Fraunhofer IOF). The air bearing consists of the
two hemispheres (shown in pale orange). The frame of the grid and the mirror
on which it is mounted by means of three micrometer screws are shown in azure
and bright blue,
respectively. {\bf Right:} Installation of PolKa in the Cassegrain cabin of the
APEX telescope. The various devices are labelled in the photo (RHWP stands for
reflecting half-waveplate).}
\label{fig:polka}
\end{figure*}

The Cassegrain focal plane of the APEX telescope is reimaged to the bolometer array of
LABOCA \citep{2009A&A...497..945S} with three concave mirrors, two flat mirrors and a lens.
The aperture ratio of the Cassegrain focus, $F/D = 8.0$, is reduced to $F/D = 1.75$ in the
focus of LABOCA. We stress that the accomodation of a polarimeter was in the design specifications of the
tertiary optics right from the beginning. In 2009, when PolKa was permanently
installed, it simply replaced one of the two flat mirrors and a filterwheel
with two analyzer grids was added.
The modulator of PolKa (Fig.~\ref{fig:polka}) was designed and manufactured by
the {\it Fraunhofer Institute for Applied Optics and Precision Engineering}
(Fraunhofer IOF, Jena, Germany). The grid for the modulator was made in the division for submillimeter
technologies at MPIfR, has an aperture of 246~mm and
is made of $20\,\mu$m thick tungsten wires \citep{2012SPIE.8452E..06S}. For the analyzer, two identical grids were
contributed by RWTH University Aachen (Germany), each of 146~mm aperture and made of gold-coated tungsten wires.
The two analyzer grids are mounted at different position angles
in order to suppress systematic effects, an equipment
feature to which we will come back in section \ref{sec:ip}. As shown in Fig.~\ref{fig:polka}
(right panel), the analyzer grid is used only in transmission.
The smooth rotation of the grid-mirror unit is ensured by an air bearing.
It consists of a rotor (pale orange in Fig.~\ref{fig:polka}) and a
stator (shown in orange). The compressed air enters through a channel machined
into the stator and flows from the center of the bearing to the outside. The
surface of the bearing is made of two hemispheres. This ensures that the
bearing works in any orientation, which for a Cassegrain cabin is evidently an
indispensable feature. We operate the air bearing at a pressure of 4~bar, which
is appropriate given the altitude of the observatory (5105~m). The position
angles are read by an encoder; in practice,
a time stamp is written for each crossing of a well defined
reference position. Interpolations between two successive reference crossings
then provide the position angles for each record in the bolometer data stream,
sampled at a speed of 1~kHz. The whole dataset (bolometer total power counts,
and the speed and position angle of the waveplate) are then downsampled to a
frequency in the range from 25 to 50~Hz and written to raw data files (in
multi-beam FITS format, \citealp{2006A&A...454L..25M}), which are subsequently
analyzed by the data processing software described in section~\ref{sec:reduction}.

\subsection{Measurement equation}
\begin{figure}
\begin{center}
\includegraphics[scale=0.7]{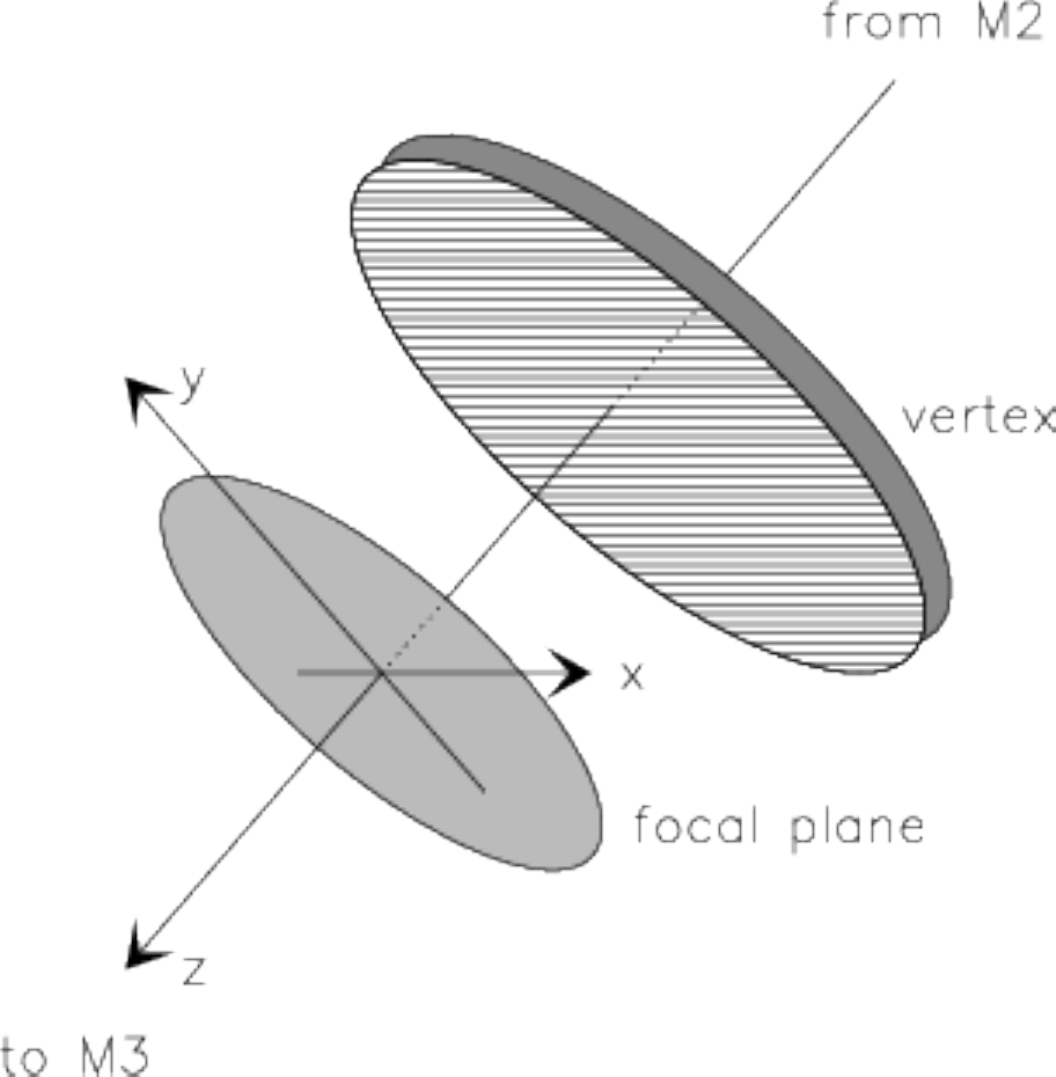}
\caption{Schematic drawing of the Cassegrain coordinate system (not to scale),
for $45\dg$ elevation. The hatched area shows the vertex, the gray area
the focal plane. The directions from the secondary mirror (M$_2$) and to the
tertiary mirror (M$_3$) are also indicated. The x-axis is parallel to the
elevation axis of the telescope. Owing to the image inversion in the focal
plane, the y-axis points towards the horizon and the x-axis to the west when
the antenna is pointed to north.}
\label{fig:focalPlane}
\end{center}
\end{figure}

The measurement equation accounts for the multiple reflections by defining a suitable coordinate frame
for the description of linear polarizations and therefore
Stokes $Q$ and $U$. The rotations and inversions of the coordinate system in the
path of the beam are then described by successive similarity transformations
of the coherency matrix \citep{1999prop.book.....B}. This algebraic formulation
of polarimetry owes its power to the capacity to represent both mixed
polarization states (like M{\"u}ller matrices) and phase information (like
complex Jones vectors). For details, we refer to appendix \ref{app:coherency}.
Here we present the result of the calculation, which is provided in full length 
in appendix~\ref{app:measurementEquation}. We define the coordinate system in
the focal plane, located between the vertex of the telescope and its tertiary
mirror, as shown in Fig.~\ref{fig:focalPlane}. In the following, we will refer
to this frame as ``Cassegrain coordinate system'' with Stokes parameters
$Q_{\rm c}, U_{\rm c}$ defined such that the polarization angle
\begin{equation}
	\psi_{\rm c} = \frac{1}{2} {\rm atan2}{(U_{\rm c},Q_{\rm c})} = \frac{1}{2}{\rm Arg}{(U_{\rm c}+iQ_{\rm c})}
\end{equation}
is measured counter-clockwise from the positive x-axis (IEEE definition),
as viewed from the tertiary to the secondary mirror.
The resulting measurement equation reads
\begin{multline}
S = \frac{1}{2} \{I-\frac{V}{2} \sin{(2\varphi-\Delta\theta)}\sin{\Phi}  \\
\tts +Q_{\rm c}\left [\cos^2{(2\varphi-\Delta\theta)}+\sin^2{(2\varphi-\Delta\theta)}
    \cos{\Phi}\right ] 
    -\frac{U_{\rm c}}{2}\sin{(4\varphi-2\Delta\theta)}(1-\cos{\Phi}) \}
\label{eq:measurement}
\end{multline}
with the Stokes parameters $I$, $Q_{\rm c}$, $U_{\rm c}$ and $V$. The signs of the 
terms for Stokes $V$ and $U$ are opposite to the sign of the Stokes $Q$ term because
PolKa is a reflecting polarimeter. $\varphi$ is the position angle of the grid but
projected onto a plane normal to the incoming ray, measured counter-clockwise
from the plane of incidence. It is related to the position angle $\varphi_0$ of the
wires of the grid (as read out by the encoder) by
$\varphi = \arctan{(\cos{\alpha}\tan{\varphi_0})}$, where $\alpha$ is the angle of
incidence defined in Fig.~\ref{fig:scheme}. Because of the relatively small angle
of incidence in our setup ($\alpha=16\fdg 23$), the difference between $\varphi$ and
$\varphi_0$ is at most $1\fdg 17$, so that the sampling of position angles is not too
distorted and remains reasonably regular. $\Phi$ is the phase difference created by
the delay between the reflected and transmitted (thus, orthogonal) polarization.
The angle $\Delta\theta$ accounts for the rotation of the $(x,y)$ plane in
Fig.~\ref{fig:focalPlane} that occurs when the beam is downfolded, and for the
orientation of the analyzer grid. More precisely, the rotation angle
$\Delta\theta$ is given by 
\begin{equation}
\Delta\theta = \theta_{\rm PA}-\theta_{\rm FP}
\end{equation}
where $\theta_{\rm PA}$ is the rotation of the beam occurring between the RPM
and the analyzer grid (measured with respect to its wires).
$\theta_{\rm FP}$ describes the beam rotation between the Cassegrain focal plane and PolKa, 
measured counter-clockwise with respect to the x-axis in the focal plane
(Fig.~\ref{fig:focalPlane}).

We deliberately show the measurement equation (\ref{eq:measurement}) in full
generality because it allows to estimate the precision needed to tune the RPM, and
to evaluate the bandwidth smearing that will be addressed below. We also keep
Stokes $V$ in equation~(\ref{eq:measurement}), although in view of the aforesaid
it can be expected to be small in
most applications. Incidentally, it may still be produced by the spurious
conversion of Stokes $I$ into Stokes $V$ arising when a grid is mounted in a
divergent beam with its wires parallel to the plane of incidence
(\citealt{1975ATTTJ..54.1665C}, \citealt{2008PASP..120..777T}). Obviously,
this occurs in every other modulation cycle. However, even if
this instrumental conversion was a first-order effect, its impact
on the measured modulation would be of second order, owing to the half-waveplate
tuning and to the relatively small phase error introduced by an imperfect tuning
and/or the bandwidth effect discussed below.

We finally note here that a fast polarization modulation may also be generated by
periodically varying the delay-induced phase difference $\Phi$. As shown by \citet{2006ApOpt..45.5107C}
and \citet{2008ApOpt..47.4429K}, a sinusoidal modulation can then be achieved by adding a
second polarization transforming reflector, rotated by
$-45\dg$ with respect to the first one (cf. Table~\ref{tab:polarimeters}). In such a design, known
as a dual {\it variable-delay polarization modulator} (dual VPM), the system can be kept free
from unwanted oscillations by using piezo-electric actuators (the {\it translational
polarization rotator} introduced by \citealp{2012ApOpt..51.6824C} is a
further development). It seems fair to say that in both approaches the stability
requirement makes the system more complex, either by adding a second VPM, or, as in our case,
an air-bearing. A way to enable an even faster modulation without moving parts is the
Faraday rotation modulator \citep{2009JPhCS.155a2006A} using a ferrite dielectric waveguide.
This kind of magneto-optical devices has a promising performance in the polarimetry of
the cosmic microwave background and of diffuse, large-scale Galactic dust emission at frequencies
up to 150~GHz \citep{2013ApJ...765...64M}. However, to implement this technology for large
detector arrays at higher frequencies is very challenging.

We now rewrite the measurement equation (\ref{eq:measurement}) for a vanishing Stokes $V$, but introduce
efficiency factors $\eta_{\rm tm}$, $\eta_{\rm bp}$ and $\eta_{\rm ts}$
describing the optical transmission, and the bandpass- and time-smearing,
respectively:
%
\begin{equation}
	S = \frac{1}{2} \eta_{\rm tm}\{I \ts + \eta_{\rm bp}\eta_{\rm ts} [Q_{\rm c}\cos{(4\varphi-2\Delta\theta)} 
		- U_{\rm c}\sin{(4\varphi-2\Delta\theta)} ]\} \tts\tts
\label{eq:measurement2}
\end{equation}
%
The efficiency factors will be quantified in the next section. A rotation of the analyzer
grid by $90\dg$ obviously inverts the signs of $Q_{\rm c}$ and $U_{\rm c}$.
By adding the signals from measurements with orthogonal analyzer orientations
it is in principle possible to separate Stokes $I$ from Stokes $Q$ and $U$. This requires a {\it strict} 
synchronization between the slew motion of the telescope and the waveplate rotation
when observing in on-the-fly mode. In practice such a synchronization is difficult to
achieve and its failure leads to a substantial loss of usable data. At best,
such an approach may work for fast on-off observations of compact sources, an
application not considered in this paper. For the extended sources
we focus on here, scanning the sky in the on-the-fly mode and
thereby obtaining a Nyquist-critical sampling, is the observing method of
choice for which a fault-tolerant system (yet accurately tracing actual values) proves to
be more efficient. We will show in section~\ref{sec:reduction} that the Stokes parameters can be
separated by a dedicated time series analysis. Notwithstanding, using two
analyzer grids at different orientations may suppress a part of the instrumental
polarization, which we will characterize below.

In order to obtain, for a given point of the mapped area, a stationary Stokes
$Q$ and $U$, we transform $(Q_{\rm c},U_{\rm c})$ to the equatorial reference
frame. From the transformation of the left-handed Cassegrain coordinates
$(x,y)$ to offsets $(\Delta\alpha\cos{\delta},\Delta\delta)$ in the
right-handed system of the tangential plane,
\begin{equation}
\left(\begin{array}{c} 
      \Delta\alpha\cos{\delta} \\
      \Delta\delta
      \end{array}
\right) = \left(\begin{array}{rr}
           \cos{\eta} & -\sin{\eta} \\
           -\sin{\eta} & -\cos{\eta}
                \end{array}
          \right)
\left(\begin{array}{c} 
      \Delta x \\
      \Delta y
      \end{array}\right)\ts ,
\end{equation}
with $\eta$ the parallactic angle, we can readily derive the corresponding
transformation of Stokes $(Q_{\rm c},U_{\rm c})$ to $(Q_{\rm eq},U_{\rm eq})$,
leading in equations
(\ref{eq:measurement}) and (\ref{eq:measurement2}) to the substitution 
$Q_{\rm c} \rightarrow Q_{\rm eq},\ts U_{\rm c} \rightarrow -U_{\rm eq},\ts
\Delta\theta \rightarrow \Delta\theta+\eta$.
In the IAU definition, which differs from the IEEE definition, the polarization
angle is measured counter-clockwise from north and is now given by
\begin{equation}
  \pa = \frac{1}{2} \left[\pi-{\rm atan2}(U_{\rm eq},Q_{\rm eq})\right]\,.
\end{equation}	
It should be kept in mind that $Q_{\rm c}$ and $U_{\rm c}$ consist of an intrinsic signal and
a polarization of instrumental origin, arising mainly in the tertiary optics.

The fractional linear polarization $\pL$ used throughout the rest of this
paper is given by
\begin{equation}
  \pL = \frac{\sqrt{Q_{\rm eq}^2+U_{\rm eq}^2}}{I}\,.
\end{equation}
Stokes $Q_{\rm eq}$ and $U_{\rm eq}$ are fraught with systematic and random
errors. The former will be removed in the correction for instrumental
polarization (see section \ref{sec:ip}), but random errors still lead to
a bias and therefore an overestimate of $p_{\rm L}$ (see e.g.,
\citealp{1974ApJ...194..249W}). However, it can be shown, both with a Monte-Carlo
simulation and analytically, that if the most likely
values of $Q_{\rm eq}$ and $U_{\rm eq}$ are used (in general median values), no
such corrections need to be done (for details see Section~\ref{sec:reduction}).
Therefore, $\pL$ and $\pa$ are calculated only at the last data reduction stage.
\subsection{Transmission properties and flux calibration}
\label{sec:transmission}
The modulation efficiency can be decomposed into the three factors
$\eta_{\rm bp},\ts \eta_{\rm ts}$ and $\eta_{\rm tm}$. From
equation~(\ref{eq:measurement2}) one can see that only the third factor affects
all Stokes parameters, while the other two factors only affect Stokes parameters
$Q_{\rm c}$ and $U_{\rm c}$.

The first factor, $\eta_{\rm bp}$ accounts for the fact that an optimal tuning
can only be achieved at a nominal frequency, while the bandpass of LABOCA,
defined by a set of cold filters at the liquid nitrogen and helium-4 stages,
extends over 60~GHz (Fig.~5 in \citealt{2009A&A...497..945S}). Even if the grid
and mirrors had a 100\% efficiency, the modulation would suffer from a loss because the optimal
half-wave phase shift ($\Phi = \pi$) can only be achieved at a fixed frequency.
Three micrometer screws allow to fine-tune the distance between the mirror and the
grid, $d({\rm M,G})$. At the passband-weighted center frequency of 344~GHz
($\lambda_0 = 870~\mu$m) and for our angle of incidence, $\alpha =16\fdg 23$,
\begin{equation}
	d({\rm M,G}) = \frac{\Phi\lambda_0}{4\pi\cos{\alpha}}
	             = \frac{\lambda_0}{4\cos{\alpha}} = 227~\mu{\rm m}\,.
\end{equation}
The tuning distance is $217\,\mu$m to account for the finite thickness of the
tungsten wires. The zero position of the micrometer screws has been confirmed by
the disappearance of the Moir{\'e} pattern arising for a finite mirror-grid distance.
At a wavelength offset $\Delta\lambda$ from $\lambda_0$, the phase is shifted from its
optimum value $\Phi = \pi$ to
\begin{equation}
\Phi = \pi+\Delta\Phi = \pi\left (1-\frac{\Delta\lambda}{\lambda_0+\Delta\lambda} \right )\,.
\end{equation}
Equation~(\ref{eq:measurement}) is then used to calculate the modulation across the
spectral bandpass. The latter may be modified by the atmospheric transmission, but
for acceptable observing conditions, the loss of modulation efficiency does not
strongly depend on the weather. For a water vapor column of 0.7~mm, $50^\circ$
elevation, and 553~hPa ambient pressure the resulting modulation efficiency is
$\eta_{\rm bp} = 99.3$\%. Because $\Delta\Phi < 0.1$~rad, the bandpass
smearing is a second order effect which explains why it affects the
polarization efficiency only weakly.

The second factor, $\eta_{\rm ts}$, results from the time smearing,
i.e., the dilution of the modulation due to the elementary integration time
step $\Delta t$, given by the sinc function
\begin{equation}
\eta_{\rm ts} = \sin{(2\omega\Delta t)}/(2\omega\Delta t)
\end{equation}
where $\omega=2\pi f_0$ is the angular speed of the waveplate rotation.
In practice, we use $f_0 = 1.56$~Hz and for integration intervals of
$\Delta t=20$ or 40~msec the dilution factors are 97 and 90\%, respectively. 

The third factor, $\eta_{\rm tm}$, can be calculated numerically from the optical properties
of the polarizer (i.e., the grid labeled $G$ in Fig.~\ref{fig:scheme}), namely its reflectance,
$R_{\|}$ and transmission, $T_{\perp}$ for the radiation power polarized parallel, 
respectively perpendicular, to the wire grids (\citealt{2012ApOpt..51..197C}, further
references therein). In the limiting case where $\lambda > 2p$, $\lambda \gg a$, and $a/p < 1/2\pi$
(where $p$ is the spacing between successive wires and $a$ their diameter) one can calculate $R_{\|}$
and $T_{\perp}$ from \citep{1898PLMS...29..523L}
%
\begin{equation}
R_{\|}= \frac{1}{1+\left( \frac{2p\cos{\alpha}}{\lambda_0}\ln{\frac{p}{\pi a}}
\right)^2}\,\,,
T_\perp= 1-R_\perp = \frac{1}{1+\left(\frac{\pi^2 a^2}{2\lambda_0 p} \right)^2}\,.
\label{eq:transmission}
\end{equation}
For PolKa, $p=(63\pm 18)\,\mu$m (the error has been determined from measurements under a microscope)
and $a=20\,\mu$m. Therefore, the filling factor $a/p$ does not obey the last of the three
conditions above. \citet{1980JPhA...13.1433C} used a Green function method to semi-analytically
derive the transmission of wire grids. Using their results \citep{1988IJIMW...9..157C} for our filling
factor of $a/p = 0.317 \pm 0.091$ and the spacing of $p/\lambda = 0.072 \pm 0.021$~$\mu$m, we
find $R_\perp = 0.1\% (+0.4, -0.1)\%$ for normal incidence, for our case of a slightly oblique incidence 
the resulting $R_\perp$ would decrease even more. We note (1) that the application of the approximative
Eq.~\ref{eq:transmission} yields a value for $R_\perp$ well within the errors due to the measured
variance of $p$, and (2) that the impact of the latter is small due to the low ratio of $p/\lambda$
(cf. Fig.~13 of \citealt{1986IMW..16...77}). \citet{2002PhD} showed with terahertz time domain
spectroscopy that the theoretical predictions from Green's function describe the actual measurements of
Tungsten wire grids similar to ours fairly well. This discussion also strongly suggests that the
polarization efficiency $\eta_{\rm bp}\eta_{\rm ts}$ in Eq.~\ref{eq:measurement2} is dominated by the
time smearing factor $\eta_{\rm ts}$.

We are therefore confident that $\eta_{\rm tm}$ is close to 0.99 (the waveplate mirror is of optical
quality and therefore leads to no loss of efficiency). Its measurement by means of primary calibrators
is notoriously difficult. When PolKa started its operation in December 2011 the flatfielding was done on
Mars without the modulator and analyzer grids.
Folding the fluxes predicted by the model of \citet{1987Icar...71..159R} (and calculated with
the online tool provided by \citealp{2008Butler}) into the bandpass of LABOCA
yields a calibration factor of 4.74~Jy/$\mu$V. The model by \citet{2006Lellouch}
yields marginally ($\sim1$\% at 300~GHz) higher continuum fluxes.
After insertion of the modulator grid calibration maps on Mars, taken without the analyzer grid,
showed no significant flux loss: During the two-week campaign, we measured daily
the flux rise of Mars (the planet approached its March 2012 opposition). The uncertainty
of the relative flux scale is typically 2\%. Comparing the model with the measured fluxes
yields a flux calibration factor of 4.63~Jy/$\mu$V. The accuracy is
limited by the uncertainty of the model flux, estimated to $\sim5$\%. -- 19 Uranus maps, 
made to characterize the instrumental polarization (see section \ref{sec:ip}), yield a calibration
factor of 4.75~Jy/$\mu$V, using the Uranus model ESA-4 (\citealp{2014Orton}, estimated to be accurate
to $2-3$\%) and assuming that the analyzer grid has a transmission of 100\% in one polarization and no
cross polarization.
The calibration factors agree within 5\% with respect to the mean value, 4.7~Jy/$\mu$V. We stress that
the observations of Uranus, unlike those of Mars, were done in polarimetry mode, which involves a more
intricate data reduction (section \ref{sec:reduction}), while the flux scale is preserved. The good
agreement between the calibrations done without grids, with the modulator grid, and then with the
analyzer grids sets a lower limit of 95\% to the transmission $\eta_{tm}$ of PolKa. 

Our calibration factor falls 27\% below the value determined by \citet{2009A&A...497..945S}. One reason for
this discrepancy is the determination of the opacity correction. We obtain it from the precipitable water vapor
($pwv$) measured once per minute with the APEX 183~GHz radiometer and converted to a
bandpass-weighted zenith opacity. A consistent and reproducible calibration is also important
in view of the processing of polarization data obtained under varying weather conditions
(see section \ref{sec:reduction}). We use the {\it am} transmission
model\footnote{Scott Paine, SMA technical memo \#152, version 7.2, February 2012}, predicting
a zenith opacity of $\tau_\nu = b_\nu(p_{\rm amb}) pwv + c_\nu(p_{\rm amb})$, where the coefficients
$b_\nu$ and $c_\nu$ parameterize the ``wet'' and ``dry'' atmosphere and depend
on frequency and outside pressure \citep{2012A&A...542L...4G}.

An accurate comparison of our calibration with that of other polarimeters is difficult. Different bolometer
arrays have different spectral bandpasses, therefore flux measurements in sources with strong spectral
indices and substantial contributions from spectral lines vary from instrument to instrument, e.g. for OMC1
(section \ref{subsec:omc1}) where spectral lines in the SCUBA bandpass contribute up to 50\% to
the observed flux \citep{1995PhDT.........3G,1999ApJ...510L..49J}.
For Tau~A (the Crab nebula, section \ref{subsec:crab}), whose
flux is dominated by synchrotron emission with a relatively flat spectral index ($S \propto \nu^{-0.3}$), our 
peak flux, applying the calibration factors derived from Mars and Uranus, differs, after correcting for the slightly
different beam sizes, by only 3.6\% from that of \citet{2004MNRAS.355.1315G}, measured with SCUBA at \wl{850}.

%
%
%
\subsection{Polarization angle calibration}
As stated in section~\ref{sec:intro}, the most important quantity for the analysis
of magnetic fields by means of polarimetry of dust or synchrotron emission
is the polarization angle. Its accuracy depends on the instrumental conversion
between Stokes $I$ on the one hand and Stokes $Q$ and $U$ on the other hand, and between
Stokes $Q$ and $U$. The first conversion can be corrected by means of an unpolarized
calibrator (we used Uranus, see section \ref{sec:ip}), while the second one
arises in the tertiary optics and can be measured with an additional polarizer.
In March 2013 we performed a series of calibrations with a high-quality grid
mounted in the focal plane of the telescope (Fig.~\ref{fig:focalPlane}), in
reflection for vertical polarization as defined in the Cassegrain reference
frame (i.e., with the wires along the y-axis of the Casssegrain coordinate system,
perpendicular to the elevation axis of the telescope).
The grid was mounted to better than $1\dg$ accuracy, and we can safely assume a
fully polarized signal with $Q_{\rm c}/I = -1$ and $U_{\rm c}/I = 0$. We
obtained $\Delta\theta = 114\fdg6$ and $2\fdg4$ for the two analyzer grids 
mounted on the filterwheel and therefore an absolute polarization angle calibration
with an accuracy well below the limitation by the sensitivity. The difference between
these observed angles is confirmed by a direct measurement to within $0\fdg2$. Our
absolute polarization angle calibration
was further confirmed by observations of celestial sources that will be presented below
together with other first light observations. The statistical error of the polarization angle
is given, for equal noise contributions from Stokes $Q$ and $U$ (which is the case here), by
$\sigma_\psi = \sigma_{\rm p_L}/2p_{\rm L}$. In astronomical polarimetry, cutoffs of 2 to $3~\sigma_{\rm p_L}$ 
are commonly used, i.e., statistical polarization angle errors of up to $14\fdg 3$ respectively $9\fdg 5$
are tolerated.
\subsection{Instrumental polarization}
\label{sec:ip}
Thanks to its axisymmetric design, the level of instrumental polarization (hereafter we
use the acronym IP) in a Cassegrain telescope is insignificant. As a matter of fact, the main
contributions to IP arise from the tertiary optics in the receiver cabin (see also
\citealp{2008PASP..120..777T} and further references therein), and ignoring
them may lead to a severe misinterpretation of polarization data. In order to quantify
and correct the IP, one ideally observes an unpolarized, unresolved source, e.g., a gas planet
like Uranus. In principle, Mars and Mercury may also be useful. Their weak, radial polarization
pattern cancels out if their disks remain unresolved (Mercury should be observed near full
phase).

\begin{figure}
\begin{center}
\includegraphics[scale=0.42]{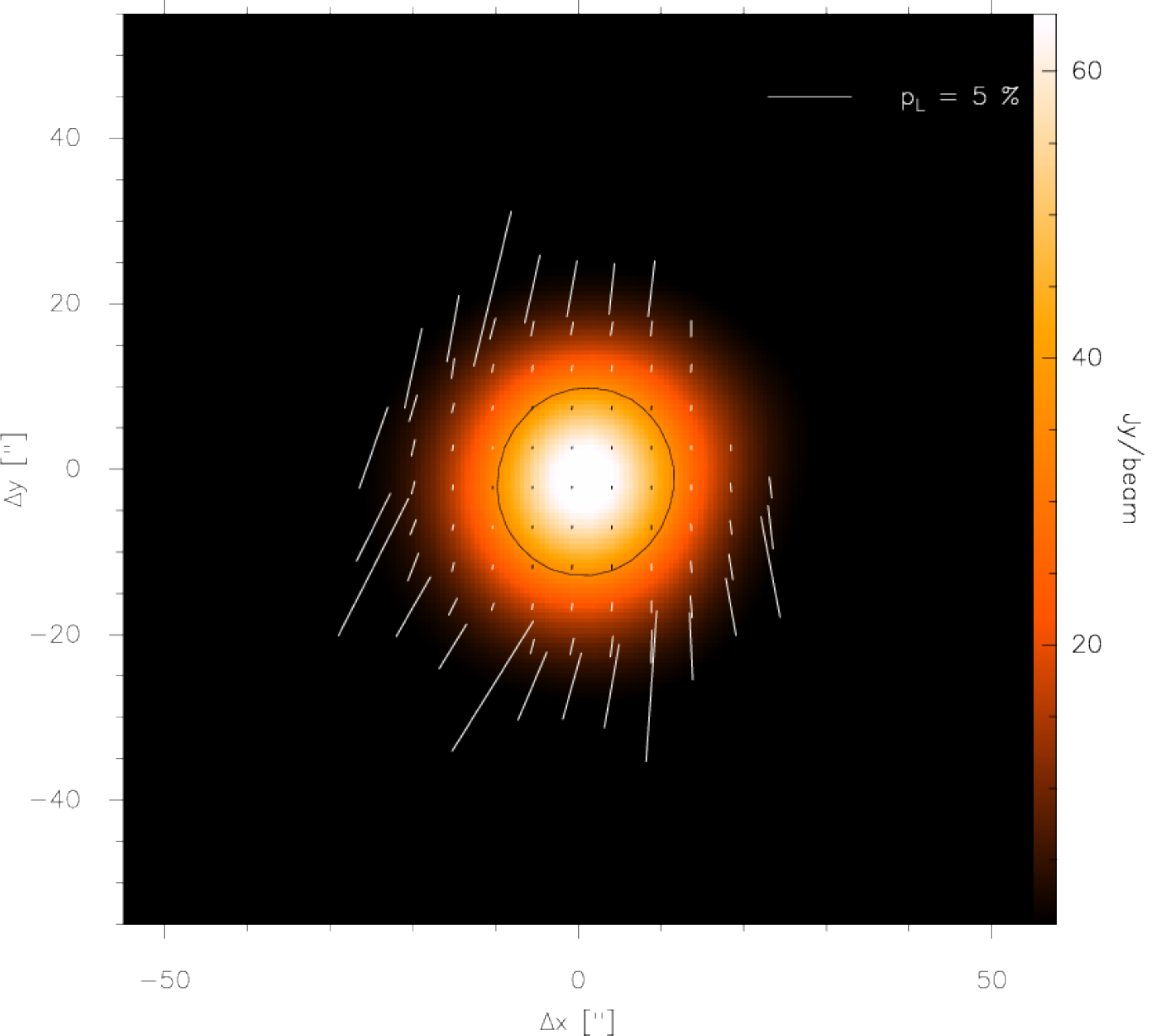} 
\caption{Stokes $I$ image of Uranus (color scale as given by wedge to the
right-hand side). Vectors of the instrumental polarization are overlaid, as
defined in the Cassegrain reference frame (counting the polarization angle ccw
from the positive x-axis, cf. Fig.~\ref{fig:focalPlane}). A polarization of
5\% is indicated in the upper right-hand corner. The black contour is at the
half-maximum level of Stokes $I$. Only polarizations with $\pL \ge 3\sigma_{\rm p_L}$
are shown.}
\label{fig:Uranus}
\end{center}
\end{figure}

Fig.~\ref{fig:Uranus} shows the IP measured in on-the-fly mode on Uranus in the Cassegrain
coordinate system. As in the map-making of Stokes I, all available LABOCA pixels have been used,
scanning the planet at different times. The resulting polarization is therefore a weighted average
of the IP at different distances from the optical axis. We expect sign changes of the instrumental
Stokes Q and U across the field-of-view; therefore the IP in the averaged data partially cancels out.
Because the sensitivity of the IP map is insufficient to be used for individual bolometer pixels, we
cannot quantify to which extent this happens, but it seems fair to conclude that the IP in
the final map, produced from the full dataset, is lower than for the individual pixels.
The fractional linear IP towards the brightness peak of the Stokes $I$ beam amounts to $p_{\rm L} = (0.10 \pm 0.04)$\%.
The spatial average within the 10~dB contour yields $(0.33 \pm 0.09)$\%. Using only the first analyzer
grid yields $(0.17 \pm 0.06)$\% at the peak position and $(0.47 \pm 0.18)$\% within the
$-10$~dB contour, and with the second analyzer we obtained ($0.09 \pm 0.04)$\% respectively $(0.37 \pm 0.36)$\%.
These numbers demonstrate that the instrumental Stokes $Q$ and $U$ beams are wider than the Stokes $I$ beam, leading
to a larger IP at off-source positions. They also show that using an analyzer grid at different orientations
can help to decrease the IP level. A further reduction occurs because the IP pattern, whose polarization vectors
are fixed in the Cassegrain coordinate system, is smeared out on the sky thanks to the parallactic rotation.
However, the above numbers substantiate that the removal of IP with a dedicated correction algorithm
is essential. In appendix \ref{app:ipcorr} we present such a procedure that accounts for the detailed
coupling of instrumental Stokes $Q$ and $U$ beam patterns to the brightness distribution on the sky.
This approach is far more sophisticated than the {\it a posteriori} application of a constant fractional IP
to the final polarization maps. An IP map like that shown in Fig.~\ref{fig:Uranus} can be used for our correction
method, provided that in the sky plane the sampling of the source resembles as closely as possible to that of the IP
calibrator. Nothing can be said about the IP below the $-10$~dB contour due to the limitation of the dynamic range.
To what extent this residual IP affects the accuracy of the measured polarization is difficult to say without
deeper observations, each at several parallactic angles. It seems fair to say that only sensitive observations
of weak polarizations ($\la 10$~mJy) are concerned when a strong source (Stokes~$I \sim 100$~Jy) is located in
the error beam.
\section{Data reduction}
\label{sec:reduction}
\begin{figure}
\begin{center}
\includegraphics[scale=0.84]{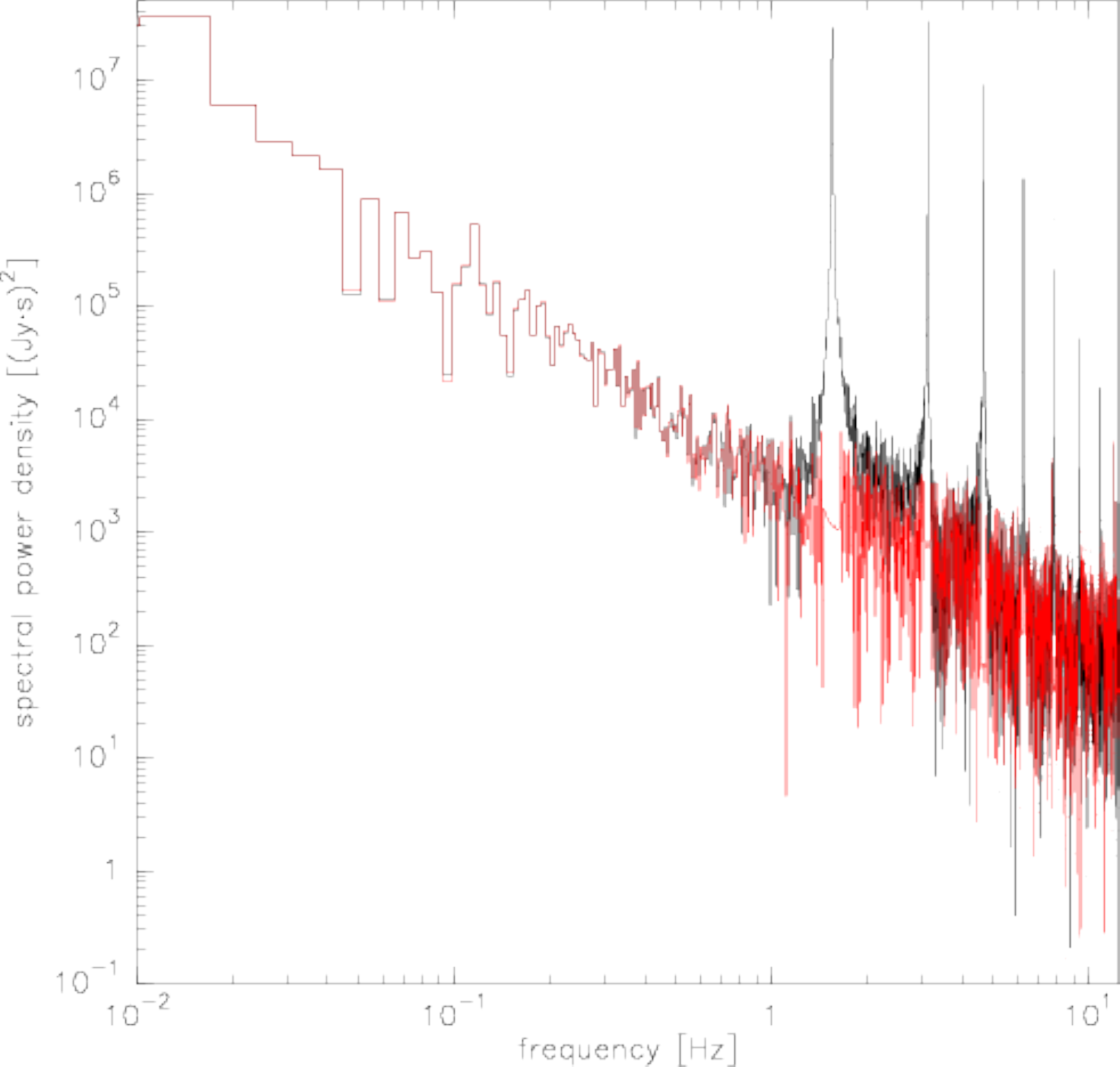}
\caption{Spectral power density of the time series of a single on-the-fly map with a spiral stroke
pattern, as received from the central pixel of the bolometer array before (black) and after
(red) removal of the beating (the spectrum does not reach a zero power density due to the 
noise bias). The 1/f noise from the atmospheric fluctuations is visible at frequencies below $\sim$3~Hz.
}
\label{fig:freqSpectrum}
\end{center}
\end{figure}

Thanks to the large field of view of LABOCA, PolKa is a polarimeter of choice
for imaging the magnetic field structure of extended objects. So far fields as large
as $10\arcmin$ have been observed; larger areas require mosaics. The observing methods
are the same as for non-polarimetric maps, i.e., a Nyquist-critical sampling is achieved by slew
motions along spiral or linear patterns \citep[see][section 8]{2009A&A...497..945S}. 
In this mode, the noise-equivalent flux density per pixel is 55~mJy$\sqrt{\rm s}$
(sensitivity weighted mean value of all usable pixels after skynoise filtering,
\citealp{2009A&A...497..945S}). However, owing to the modulation of the signal, the
data reduction methods are more involved. PolKa is usually operated at a spinning
frequency of $f_0 = 1.56$~Hz, or a modulation frequency of
$f_{\rm M} = 4 f_0 = 6.24$~Hz, so as to obtain four data records of 40~msec per
modulation cycle. This frequency also allows to separate the modulation of the
polarized flux from atmospheric fluctuations which remain below $\la 3$~Hz, as
demonstrated by the spectral power density (i.e., the Fourier transform of the
autocorrelation function of the time series) shown in
Fig.~\ref{fig:freqSpectrum} (the apparent harmonic signal will be discussed in
section \ref{sec:beating}). 

A typical time series of PolKa data is shown in Fig.~\ref{fig:dataReduction}
and demonstrates the data reduction steps that will be discussed in the following.
The signal can be decomposed into three contributions, namely into (1) a periodic,
deterministic signal of instrumental origin, (2) a piecewise periodic, deterministic
signal (the polarization received from the observed source), and (3) a non-deterministic
signal (the 1/f noise from the atmosphere, the detector noise and the high-frequency
noise from the readout electronics). 
\subsection{Speed considerations}
\label{sec:speed}
Before we present the further data reduction methods, a few words about speed considerations
seem appropriate here. The lowest frequencies of the atmospheric fluctuations in
Fig.~\ref{fig:freqSpectrum} can be suppressed by choosing an adequate mapping speed. In the logarithmic spiral mode,
a single subscan takes 36~s. Atmospheric fluctuations on longer timescales are therefore avoided.
To what extent faster fluctuations degrade the image quality depends both on the source structure
and on how it is sampled in the on-the-fly observing mode, scanning the source with a linear or
spiral stroke pattern. The latter is used with a constant angular speed of $90\dg/{\rm s}$, i.e.,
a filamentary source, repeatedly appearing in the time series of a scan, has Fourier components
at a frequency of 0.25~Hz and its harmonics. The removal of atmospheric fluctations is
therefore mandatory and will be discussed in section~\ref{sec:filter}. On the other hand,
for a speed of $200''/{\rm s}$ (the largest speed occurring in the spiral stroke pattern), the main lobe
of the diffraction pattern of a point-like source has a width of 8.8~Hz ({\sc FWHM}) in frequency
domain, which implies that the largest power is at frequencies above those of the fluctuations.

These considerations hold for Stokes~$I$. Assuming a constant fractional polarization across the source,
its modulation appears in frequency domain as a scaled version of the Stokes~$I$ spectrum but now
centered at the modulation frequency $f_{\rm M}$, i.e., well above the atmospheric fluctuations,
which explains why they affect the polarization far less than Stokes~$I$. The consequences for
the demodulation will be treated in section~\ref{sec:demodulation}.
\subsection{Removal of the total power beating}
\label{sec:beating}
\begin{figure}
\begin{center}
\includegraphics[scale=0.64]{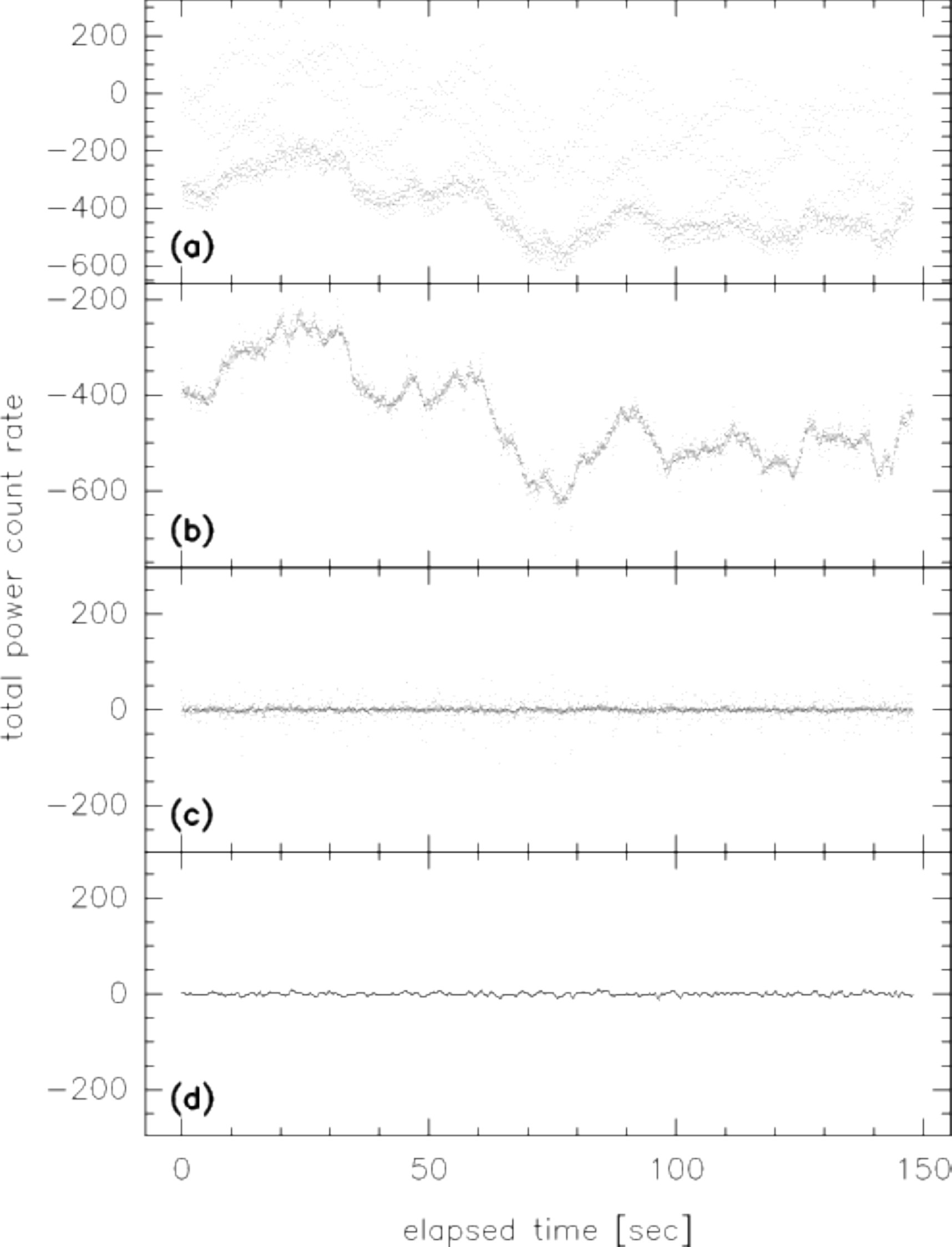}
\caption{Demonstration of the data reduction steps. From top to bottom: {\bf(a)}
Time series of raw data from the central LABOCA pixel, for a single on-the-fly map with spiral
stroke pattern. {\bf(b)} Same after removal of total power beating, {\bf(c)} after removal of
correlated noise, {\bf(d)} after removal of high-frequency noise (wavelet filter, see
appendix \ref{app:wavelet}). The vertical scale in {\bf(b)}-{\bf(d)} is fixed so
as to show the noise suppression.}
\label{fig:dataReduction}
\end{center}
\end{figure}

A closer inspection of the time series shows a spurious signal on top of the 
expected output (Fig.~\ref{fig:dataReduction}a). The spectral energy distribution
(Fig.~\ref{fig:freqSpectrum}) confirms that it is a harmonic total power beating,
starting at the fundamental frequency $f_0$ of the mechanical modulator rotation
and visible in all the harmonics up to the Nyquist frequency. The spurious signal is not due to a gain
variation and therefore independent from the total power received from the
sky. Its origin can be manifold. It is common wisdom that a modulator housed
in the cold part of the optics provides intrinsically more stable signals, but
instability is not an issue here (the beating is a deterministic signal). The
resonances that can occur in reflecting polarizers \citep{2001PASP..113..622H,2008ApOpt..47.4429K}
are also unlikely to be of concern here (the characteristics of the grid as described
in section~\ref{sec:transmission} are optimal).

Imaging the beating across the full array and as a function of time, i.e., record by record, reveals a 
bar-like, asymmetric total power distribution, rotating about the center of the array with the frequency $f_0$.
The strength of this rotating feature is modulated with the frequency $2f_0$ of the
second harmonic. In the time series of the total power beating the combination of these spatial and temporal
variations leads, for a given bolometer channel, to the observed profile. We also note here that the sampling frequency
of 25~Hz is not an exact multiple of $f_0$ and that the spikes of the beating are not fully resolved
in time; in the Fourier spectrum this leads to a further redistribution of spectral power among the harmonics. 
This discussion may suggest that the beating is due to the differential emissivity of the grid and the mirror of the modulator,
so that the higher emissivity of the mirror leads to a net polarization
perpendicular to the wires of the modulator grid. However, we cannot corroborate such a conclusion.
The mirror is made from an aluminium alloy whose conductivity falls short of that of pure aluminium but
can be exptected to be of the order of $\sigma = 2.5\times 10^7$~S/m, while that of tungsten is $1.8\times 10^7$~S/m.
From the Hagen-Rubens law for the spectral emissivity, $\epsilon_\nu = 4\sqrt{\pi\nu\epsilon_0/\sigma}$, and
Kirchhoff's law we can then determine the expected emission. Assuming for the modulator a typical temperature of
283~K, we obtain a polarized signal with a Rayleigh-Jeans temperature of $\sim 200$~mK, accounting for the filling
factor of the grid. The strength of the total power beating is comparable to that of our Orion~KL data, i.e.,
a $\sim 100$~Jy source observed with an atmospheric transmission of 55 to 72\%. With an aperture efficiency of
0.6 \citep{2006A&A...454L..13G}, this converts to an antenna temperature of 1.4 to 1.8~K which is
an order of magnitude above our estimate of the beating expected from a differential emissivity
of mirror and grid. More dedicated measurements will be needed to clarify the origin of the beating and
to further improve the system.

\begin{figure}
\begin{center}
\includegraphics[scale=0.60]{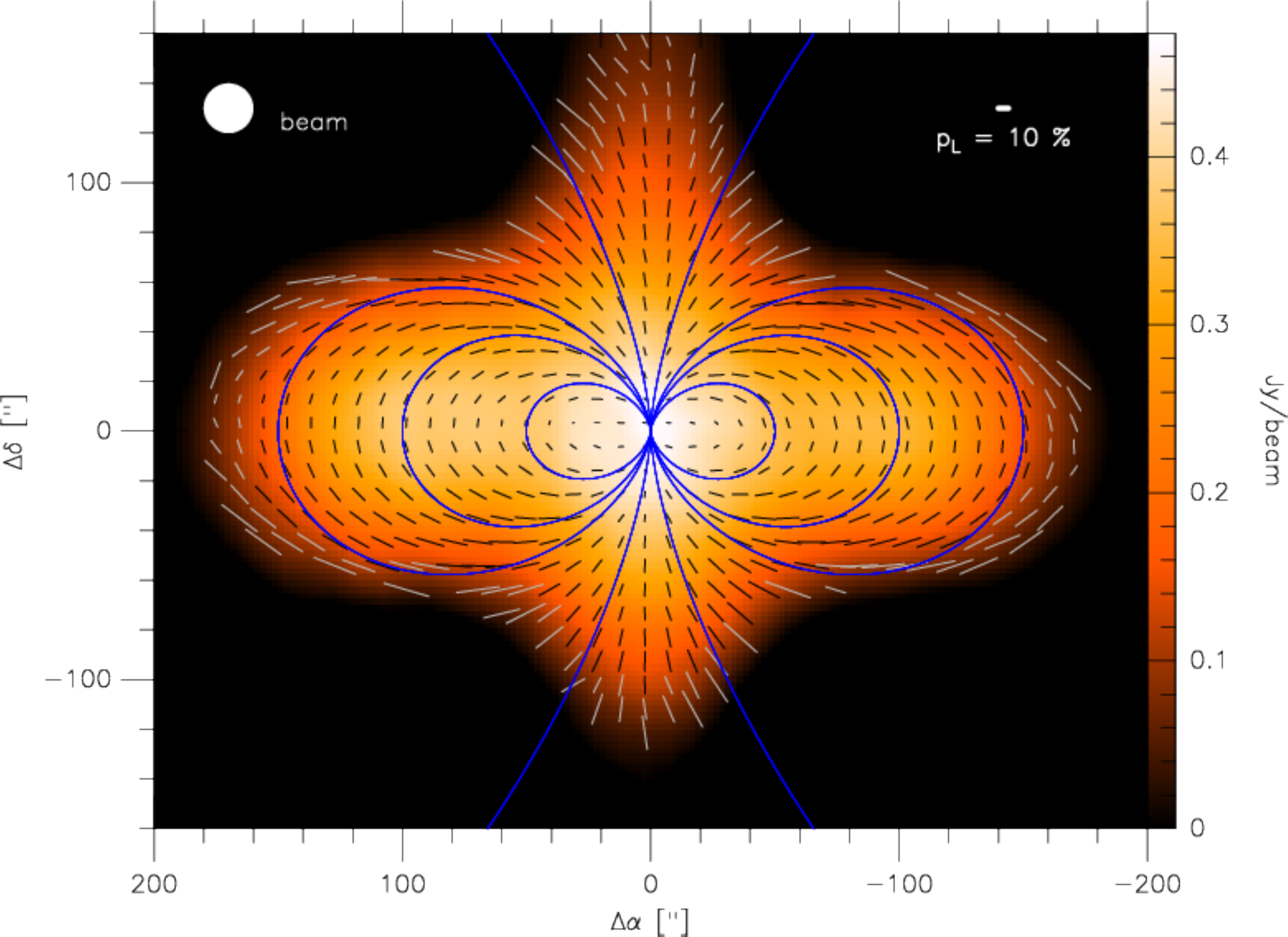}
\caption{Reconstruction of a test source with linear polarization along dipole field lines
(blue contours, $p_{\rm L}=10$\%). The linear polarization deduced by the data reduction is
shown as black or white vectors. The emission in Stokes $I$
(color scale) consists of three Gaussian brightness distributions. The simulation includes
atmospheric total power fluctuations and a total power beating. 
The on-the-fly sampling of this test source is the same as for Fig.~\ref{fig:crab}. Polarization vectors
are shown for a Stokes $I$ emission above 1~mJy. For details see section~\ref{sec:beating}. The beam size
($20\arcsec$ {\sc fwhm}) and a 10\% linear polarization are shown in the upper left and right
corners, respectively.}
\label{fig:model}
\end{center}
\end{figure}

In practice, the total power beating dominates the signal but is strictly harmonic.
In the raw data from a single scan and pixel, the
modulation of the signal due to its polarization is still negligible; with the
aforementioned noise-equivalent flux density, the noise in a 40~msec dump amounts
to 275~mJy. 
In the Fourier transform of the time series the beating can be easily distinguished from the
spectrum of white noise (which has a constant amplitude but a random phase) and 
is removed from the time series pixel-wise and scan by scan
(Fig.~\ref{fig:dataReduction}b), by eliminating the narrow spikes in the Fourier spectrum of the 
adequately apodized time series, and interpolating between the real and imaginary parts next to them.
In practice, this correction leads to an
insignificant distortion of the observed polarization, because the latter is only 
piecewise periodic. Its spectral power is distributed across a much wider frequency range
determined by the window function (unity when a pixel crosses the source, and zero elsewhere),
especially if a filamentary structure is observed perpendicular to its long axis. Another reason for
the widened Fourier spectrum of the modulation centered at $4f_0$ is that the polarization vectors on
the sky, while the source is scanned, either rotate with, or counter-rotate against, the motion of the
waveplate. Tests with simulated sources have shown that the removal of the beating
does not significantly affect the measurement of the intrinsic linear polarization. One of these
test sources, a $6\arcmin$ wide model consisting of three Gaussians, whose power is as weak as 0.5\% of that
of the simulated beating (100~Jy), is shown in Fig.~\ref{fig:model}. After removal of the beating 
and of the simulated atmospheric total power fluctuations, the intrinsic polarization of
$\pL=10$\% (modeled as the projection of a dipole field onto the plane of the sky)
could be restored. Towards zones with a strong curvature of the field lines the
$20''$ wide beam ({\sc fwhm}) leads to depolarization, while an excess of the
fractional polarization is observed where the Stokes $I$ emission is weak; this is
due to the difficulties to conserve the largest spatial scales in the reconstruction 
of the data. In summary, the removal of the total power beating by the software filter
described in this section shows that mounting a polarimeter in the warm part of the optics, making
it more prone to resonances of this kind, can be compensated for by an adequate data reduction
algorithm.
\subsection{Time series filtering}
\label{sec:filter}
The overlap of the near-field beam patterns of the individual pixels leads to
a strong correlation between the signals received by a pair of pixels, because
they ``see'' the same atmospheric fluctuations ("1/f noise"). For each data record the median signal across
the array provides a good measure for the dominating contribution of the 
atmosphere to the received total power. However, the gains of the individual
bolometer pixels, as determined from observations of Mars, Uranus or Neptune,
are not appropriate because, unlike these primary calibrators, the atmospheric 
emission fills the entire forward-beam power pattern of each
pixel. The strong correlation among the signals allows us to re-calculate and
to apply the gains for the reception of the atmospheric total power. Then for
each record the median signal across the array is calculated and removed from
the time series, and the resulting signal is scaled to the correct far-field
gain for each pixel. 

While the 1/f noise is efficiently suppressed by means of this correlated-noise
filter, the power at the highest frequencies ($f > 10$~Hz) originates from the readout
electronics and is suppressed by means of a wavelet filter. We refer the reader interested
in wavelet filtering to appendix \ref{app:wavelet}.

Both the correlated-noise and wavelet filtere may not be used for the modulated (i.e., polarized) part of the
signal. Depending on the polarization structure of the observed source, the
modulation leads to a partial correlation of the signals and would be distorted
by the removal of correlated noise. Fortunately, such a step is not necessary here:
As demonstrated in Fig.~\ref{fig:freqSpectrum}, the atmospheric fluctuations hardly
leak into the modulation of the polarized fraction of the total power. We note, however,
that although it is not modulated Stokes $I$ contaminates the polarization, due to the scanning
motion. This happens, e.g., for a point source scanned at maximum speed, leading to an
8.8~Hz wide ({\sc fwhm}) Fourier spectrum (the most extreme case demonstrated in section~\ref{sec:speed}).
The correction for this contribution in the demodulation of the polarization will be described
in section~\ref{sec:demodulation}.

For the modulated, i.e., polarized fraction of the signal a frequency filter of the form
\begin{equation}
g(f) = \frac{1}{2}\left (1-\cos{\frac{2\pi(f-f_{\rm low})}{f_{\rm high}-f_{\rm low}}}\right) \mathrm{\,\,for\,\,} f_{\rm low} \le f \le f_{\rm high},\,\, g(f) = 0 \mathrm{\,\, elsewhere}
\end{equation}
is applied to the Fourier transform of suitably apodized segments of data 
(typically individual scans), with $f_{\rm low} = 3$~Hz and $f_{\rm high} = 10$~Hz.

\subsection{Map-making and demodulation}
\label{sec:demodulation}
Depending on the polarization structure of the source, the phase and amplitude of 
the modulation changes rapidly when the telescope sweeps across the sky. Therefore,
the demodulation of the data and the map-making must be performed by
the same data reduction step, to retrieve Stokes $Q_{\rm eq}$ and
$U_{\rm eq}$. The map-making, i.e., the gridding of Stokes $I$ to a regular
grid, follows the usual
procedure to construct a regularly sampled image from a critically, but
irregularly sampled stream of data: For each pixel of the output image,
only data within a cutoff radius around this pixel will be considered, and the
flux assigned to this pixel is an average of this data, weighted with a
convolution kernel (here a Gaussian is used). For Stokes $Q_{\rm eq}$ and $U_{\rm eq}$, the approach
is different due to the modulation: first, the set of data located within the cutoff radius is
binned into discrete waveplate position angle intervals, then this binned data is 
demodulated. 
In the logarithmic spiral mode \citep{2009A&A...497..945S}, LABOCA scans the sky at a varying speed (at
constant angular velocity). Therefore the sampling of waveplate angles is no longer
strictly synchronized with the spatial sampling of the area to be mapped, and the
demodulation scheme needs to be
generalized\footnote{Even for the case of a regularly sampled map observed with
a waveplate rotating at constant speed, for a non-zero angle of incidence the sampling
of position angles cannot be strictly regular due to the projection effect.}.
We will show now that this generalization is straightforward.

The measurement process can be represented via equation~(\ref{eq:measurement})
as a time series, whose elements correspond to discrete time steps and
therefore
different waveplate angles and celestial positions. We write this time series as
a vector $\bm{\mathcal A}$. Since the polarization angle changes when the
telescope sweeps the source, for the demodulation process we have to consider
a subset of the data, $\bm{\mathcal B} \subseteq \bm{\mathcal A}$, whose
distance from a given pixel of the output map is within the cutoff radius of
the convolution kernel used in the map-making. As already mentioned,
$\bm{\mathcal B}$ may still contain a residual of the unmodulated signal, i.e., 
Stokes $I$. This residual is removed from $\bm{\mathcal B}$ by a 
baseline subtraction, and we are left with the modulated, i.e., polarized,
signal fraction only, $\bm{\mathcal B}_{\rm m}$. The aim of generalized
synchronous demodulation is to construct weight vectors $\bm{w}_{\rm Q}$,
$\bm{w}_{\rm U}$, such that Stokes $Q$ and $U$ are obtained through the scalar
products
\begin{equation}
Q = \bm{\mathcal B}_{\rm m} \cdot \bm{w}_{\rm Q},\,\,
U = \bm{\mathcal B}_{\rm m} \cdot \bm{w}_{\rm U}\,.
\label{eq:demodulation}
\end{equation}
In the following we assume that PolKa is at its nominal $\lambda/2$ tuning,
and denote the corresponding sine and cosine time series as vectors with
elements
\begin{equation}
	{\mathcal C}_{\rm j} := \cos{4\varphi_{\rm j}}, \,\,
	{\mathcal S}_{\rm j} := \sin{4\varphi_{\rm j}} 
\label{eq:sine}
\end{equation}
where $\varphi_{\rm j} = \varphi(t_{\rm j})$.
The sensitivity of Stokes $Q$ and $U$ can then be derived from the radiometric
noise $\sigma_{\rm rms}$ in the time series $\bm{\mathcal A}$ with 
\begin{equation}
\sigma_{\rm Q,U}=\sigma_{\rm rms}\sqrt{\bm{w}_{\rm Q,U}\cdot\bm{w}_{\rm Q,U}}.
\end{equation}
Because the sampling is in general irregular, we cannot expect that
$\bm{\mathcal C}\cdot\bm{\mathcal S} = 0$ (which was the prerequisite for
the demodulation scheme in \citealt{2004A&A...422..751S}). However, generalized
synchronous demodulation offers many possibilities for the construction of
the weight vectors $\bm{w}_{\rm Q}$, $\bm{w}_{\rm U}$, also for other than
sinusoidal modulations. In our case an obvious choice consists
of using $\cos{4\varphi}$ and $\sin{4\varphi}$ as basis functions,
\begin{equation}
   \bm{w}_{\rm Q} = \mu \bm{\mathcal{C}} + \nu \bm{\mathcal{S}},\,\,
   \bm{w}_{\rm U} = \xi \bm{\mathcal{C}} + \rho \bm{\mathcal{S}}
\end{equation}
and to determine the coefficients $\mu$, $\nu$, $\xi$ and $\rho$ such that
equations~(\ref{eq:demodulation}) are fulfilled. This yields
\begin{equation}
	\mu = \frac{\bm{\mathcal S}^2}
                   {\bm{\mathcal C}^2 \bm{\mathcal S}^2
                    -\left(\bm{\mathcal S}\cdot\bm{\mathcal C} \right)^2
	           },\,\,\,\,
        \nu = \xi = \frac{-\bm{\mathcal S}\cdot \bm{\mathcal C}}
                   {\bm{\mathcal C}^2 \bm{\mathcal S}^2
                    -\left(\bm{\mathcal S}\cdot\bm{\mathcal C} \right)^2
		   },\,\,\,\, \nonumber \\
        \xi = \nu,\,\,\,\,
        \rho = \frac{\bm{\mathcal C}^2}
                    {\bm{\mathcal C}^2 \bm{\mathcal S}^2
                     -\left(\bm{\mathcal S}\cdot\bm{\mathcal C} \right)^2
	            }       
\label{eq:coefficients}
\end{equation}
%
%
Thanks to the Cauchy-Schwarz inequality the denominator of these coefficients
is positive, and zero only in the case that $\bm{\mathcal C}$ and
$\bm{\mathcal S}$ are linearly dependent, i.e., for $\tan{4\varphi_{\rm j}} = 1$ or
$-1$ for all $\varphi_{\rm j}$. The separation of Stokes $Q$ from Stokes $U$ in a
technique applying a sinusoidal modulation is then impossible, but this situation
is unlikely to occur, not least due to the parallactic rotation.

\section{First light observations}
\label{sec:results}
\subsection{The Moon}
\begin{figure}
\begin{center}
\includegraphics[scale=0.90,angle=-90]{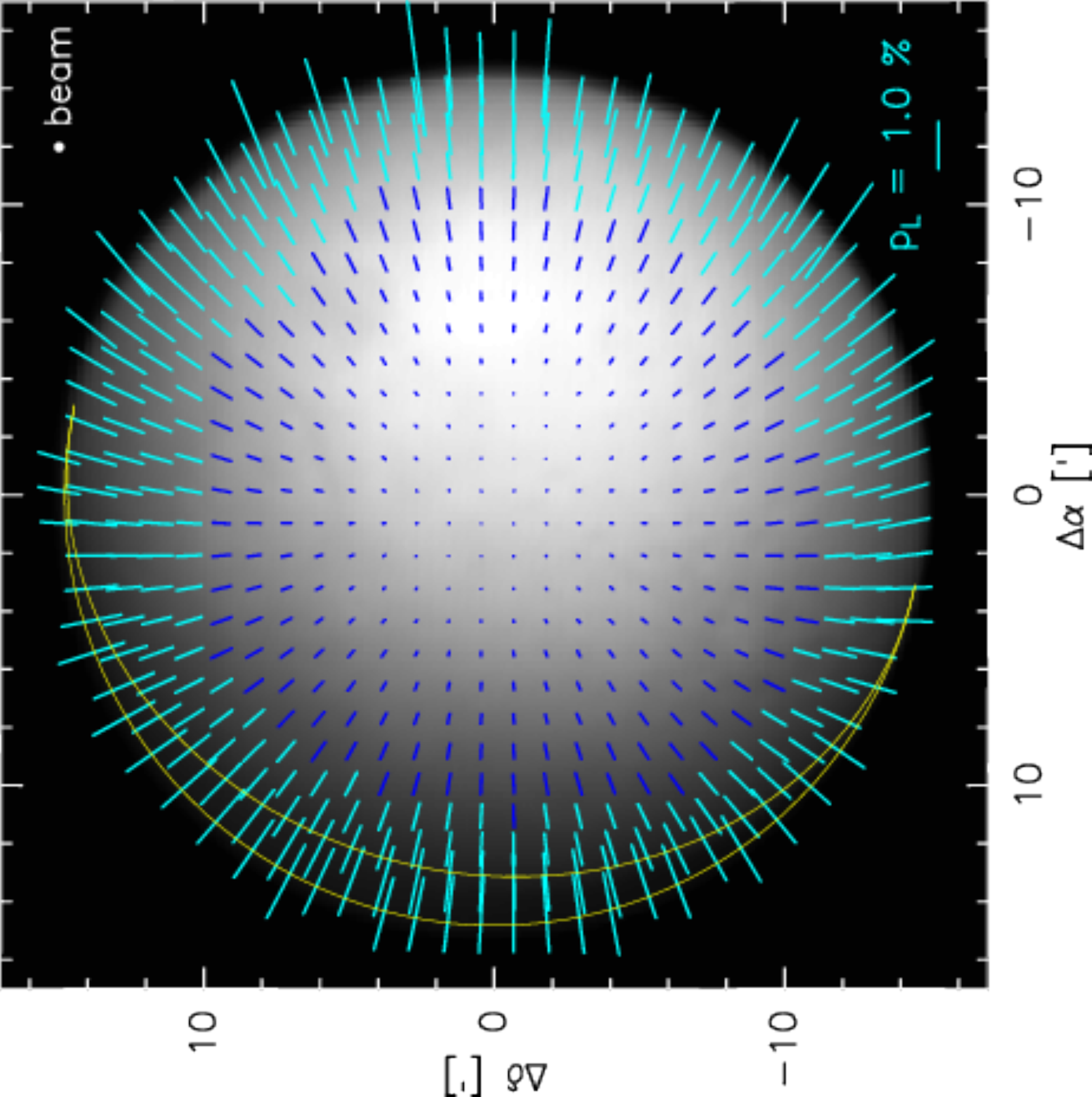}
\caption{Linear polarization of the lunar submillimeter emission. A fractional
linear polarization of 1\% is indicated in the lower right corner. The yellow
contour shows the position of the terminator. The PolKa beam ($20\arcsec$
{\sc fwhm}) is indicated in the upper right corner.}
\label{fig:moon}
\end{center}
\end{figure}

The lunar continuum radiation from radio to far infrared wavelengths is
dominated by the thermal emission from the regolith covering the surface,
originating from a frequency dependent depth of 10~m at 3~GHz (the typical
thickness of the regolith layer, \citealp{1975Icar...24..211K}) to a few
centimeters at 37~GHz \citep{2007Icar..190...15F}.
The Fresnel coefficients for the last refraction between the regolith and the 
vacuum are different for polarizations perpendicular and parallel to the plane
of incidence. The resulting radial linear polarization is a useful calibration
source for polarimeters. As part of the commissioning of PolKa, the Moon was observed on
2011 Dec 8, at phase 94.1\% shortly before full moon, scanning the disk in
zig-zag mode twice, using different position angles for the filterwheel. The
expected radial polarization pattern has been reproduced, and the fractional
linear polarization amounts to up to $\sim2$\% (Fig.~\ref{fig:moon}). The
terminator is not sharp, due to the delayed heating of the subsurface layers.
The coupling of the polarized sidelobes due to the telescope's error beam
pattern, irrelevant for the observations of more compact structures, leads to a 
slight deviation from a purely radial pattern, owing to the phase effect. 
\subsection{Tau~A}
\label{subsec:crab}
\begin{figure}
\includegraphics[scale=0.35,angle=0]{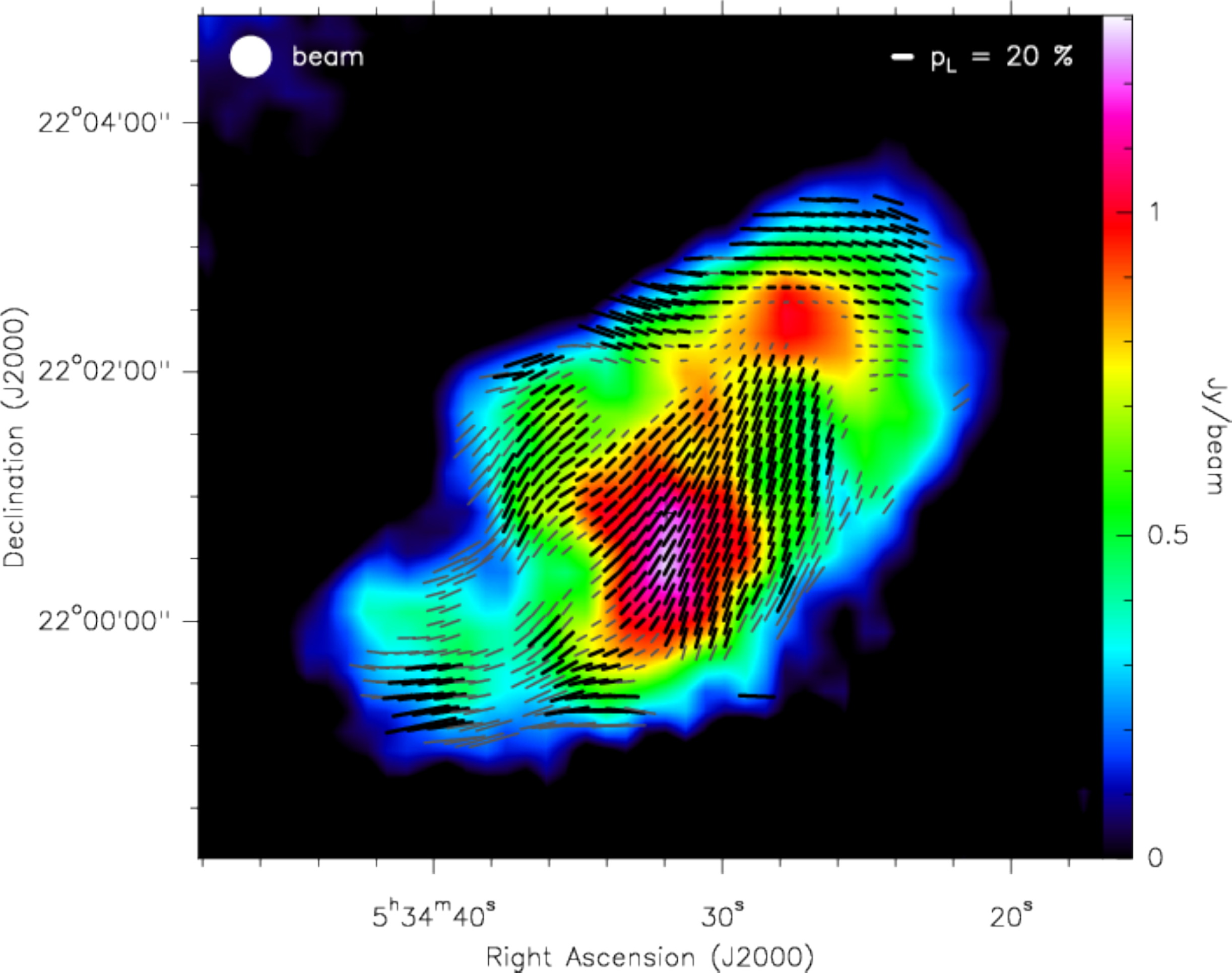} 
\includegraphics[scale=0.35,angle=0]{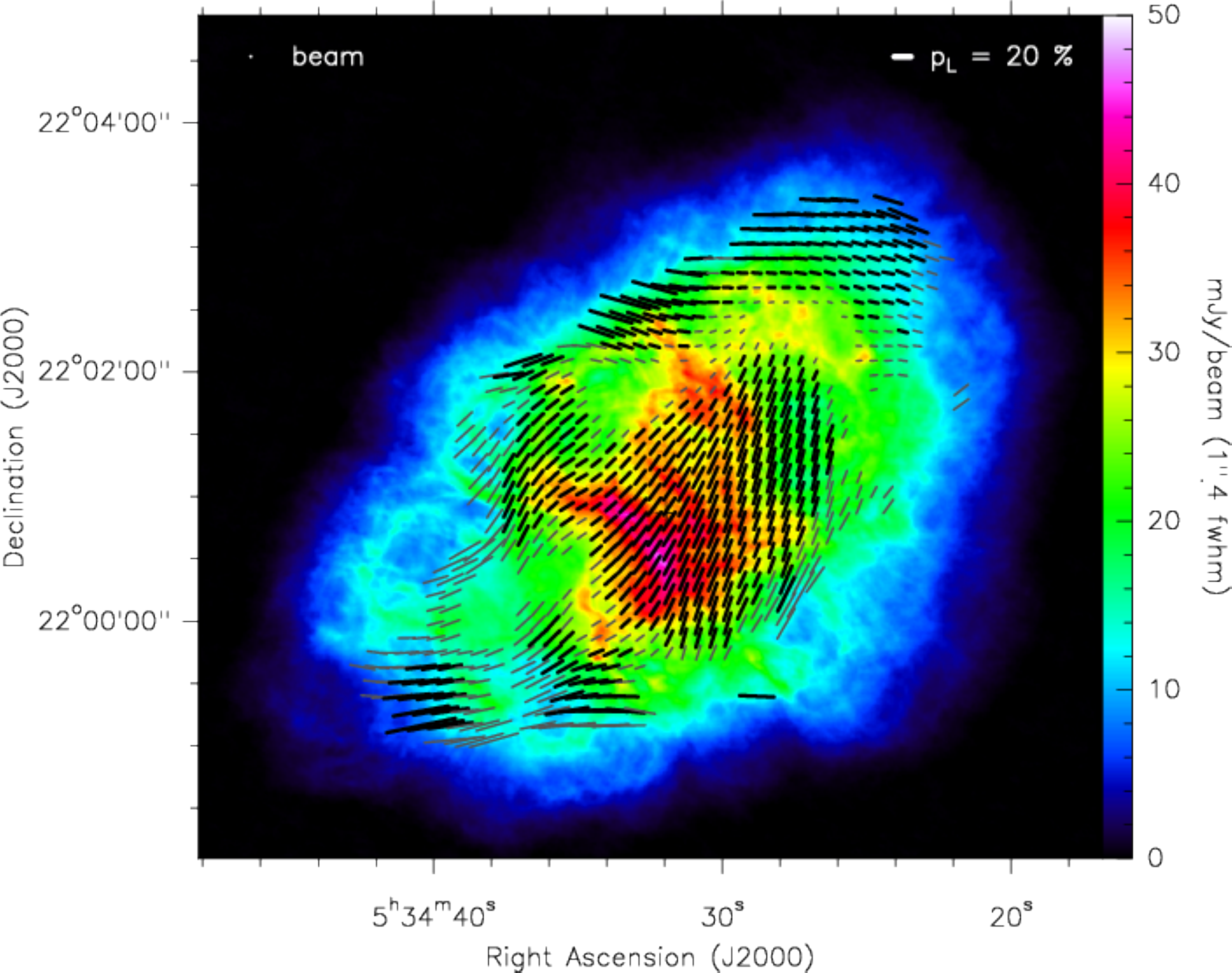} 
\begin{center}
\caption{\wl{870} polarization vectors (corrected for instrumental polarization)
in Tau~A. Polarizations above $3\sigma_{\rm p_L}$ (i.e., $\sigma_\psi \le 9\fdg 5$)
are shown in black, those with $2\sigma_{\rm p_L} \le p_{\rm L} < 3\sigma_{\rm p_L}$
in gray ($\sigma_\psi \le 14\fdg 3$). A linear polarization of 20\% is indicated in the top
right corners. The black cross marks the pulsar position.
{\bf Left:} with Stokes \wl{870} Stokes I emission underneath (plot scale to the right).
The $20''$ ({\sc fwhm}) beam is shown in the upper left corner. {\bf Right:}
with VLA 5~GHz continuum ($1\farcs 4$~{\sc fwhm}, archive data, 
\citealp{2001ApJ...560..254B}).}
\label{fig:crab}
\end{center}
\end{figure}

Tau~A, the Crab nebula, is a plerion-type supernova remnant. The polarization of
its radio emission, discovered in 1957 independently by \citet{1957ApJ...126..468M}
and \citet{1959SvA.....3...39K}, has confirmed synchrotron radiation as the underlying
mechanism. The distribution of the linear polarization across the center of the nebula,
close to the pulsar which powers the synchrotron emission, is fairly smooth and the
polarization angle does not vary significantly from the radio emission over visible light
\citep{1971Natur.229...39F} to X-rays \citep{1978ApJ...220L.117W}. For more recent
work we refer to \citet{2008ARA&A..46..127H}. While only upper limits have been reported for the circular
polarization (at $\lambda$3~mm, $<0.2$\%, \citealp{{2011A&A...528A..11W}},
further references therein), the linear polarization at $\lambda$3~mm amounts
to up to 30\% \citep{2010A&A...514A..70A}. Our polarization map of Tau~A 
is shown in Fig.~\ref{fig:crab}; the results are summarized in Table~\ref{tab:results}
with and, for comparison, without the correction for instrumental polarization.
The map was processed from 36 on-the fly maps of 150~sec each, corresponding to an on-source
observing time of 1.5~h, at a typical atmospheric transmission of 70\%. The sensitivity
is $\sim$20~mJy/beam. The correction for instrumental polarization (see appendix \ref{app:ipcorr})
changes the linear polarization by a few percent and the polarization angle by
up to a few degrees. The overall agreement with other
polarimetry campaigns (at $\lambda$3~mm with XPOL at the 30m telescope, \citealp{2010A&A...514A..70A},
and at \wl{850} with SCUPOL at the JCMT, \citealp{2009ApJS..182..143M}) is reasonably good
towards the synchrotron emission peak. The largest discrepancies occur towards the pulsar position; for
the fractional polarization they are significant, but not for the polarization angles.
To date it is impossible to say whether this difference is intrinsic or due to a bias introduced by
the measurements and their analysis. Depolarization can be ruled out as an explanation, since
SCUPOL and XPOL measure the same fractional polarization, despite their different beams
($20\arcsec$ and $27\arcsec$ {\sc fwhm}, respectively). We note that the Pulsar position is 
$0\farcm 5$ north of the brightness peak. As mentioned in Section~\ref{sec:ip}, the instrumental
polarization may matter; the three polarimeters account for it in different ways:
\citet{2010A&A...514A..70A} applied jackknife tests to ascertain the robustness of their
results, while SCUPOL applies an approximative pixel-wise correction \citep{2003MNRAS.340..353G}. For PolKa we
refer to appendix~\ref{sec:ip}.

Here we eventually examine whether Faraday rotation or a dust contribution to the continuum emission
can rotate the polarization vector. The large-scale rotation measure towards Tau~A is $\sim-21$~rad~m$^{-2}$
\citep{1991ApJ...368..231B} and mostly external to the nebula, while in unresolved filaments of thermal gas
it rises up to $~300$~rad~m$^{-2}$. Even in the latter medium, the differential Faraday rotation between
$\lambda$3mm and \wl{850} is not measurable. --
A dust emission component of polarized flux $P_{\rm dust}$ and polarization angle
$\psi_{\rm dust}$ that adds to the synchrotron emission of polarized flux
$P_{\rm sync}$ and polarization angle $\psi_{\rm sync}$ leads to a 
rotation $\Delta\psi$ of the polarization vector, to first order in
$P_{\rm dust}/P_{\rm sync} \ll 1$, by
\begin{equation}
\Delta\psi = \frac{P_{\rm dust}}{2P_{\rm sync}}\sin \left [ 2(\psi_{\rm dust}-\psi_{\rm sync}) \right ]\,.
\end{equation}
The largest rotation of the polarization angle occurs for dust emission polarized at $45\dg$ from the
synchrotron emission. Then the $\lambda3$\,mm and \wl{870} polarization
angle difference of $15\dg$ would require the dust emission to be polarized with 
$0.52~P_{\rm sync}$ which is certainly not conceivable: \citet{2004MNRAS.355.1315G} find only a
small amount ($\la 0.07~M_\odot$, $\sim1.5$\% of the nebula mass, \citealp{2001ApJ...560..254B})
of silicate or graphite dust at a temperature of about 50~K. 

\begin{figure}
\begin{center}
\includegraphics[scale=0.35,angle=0]{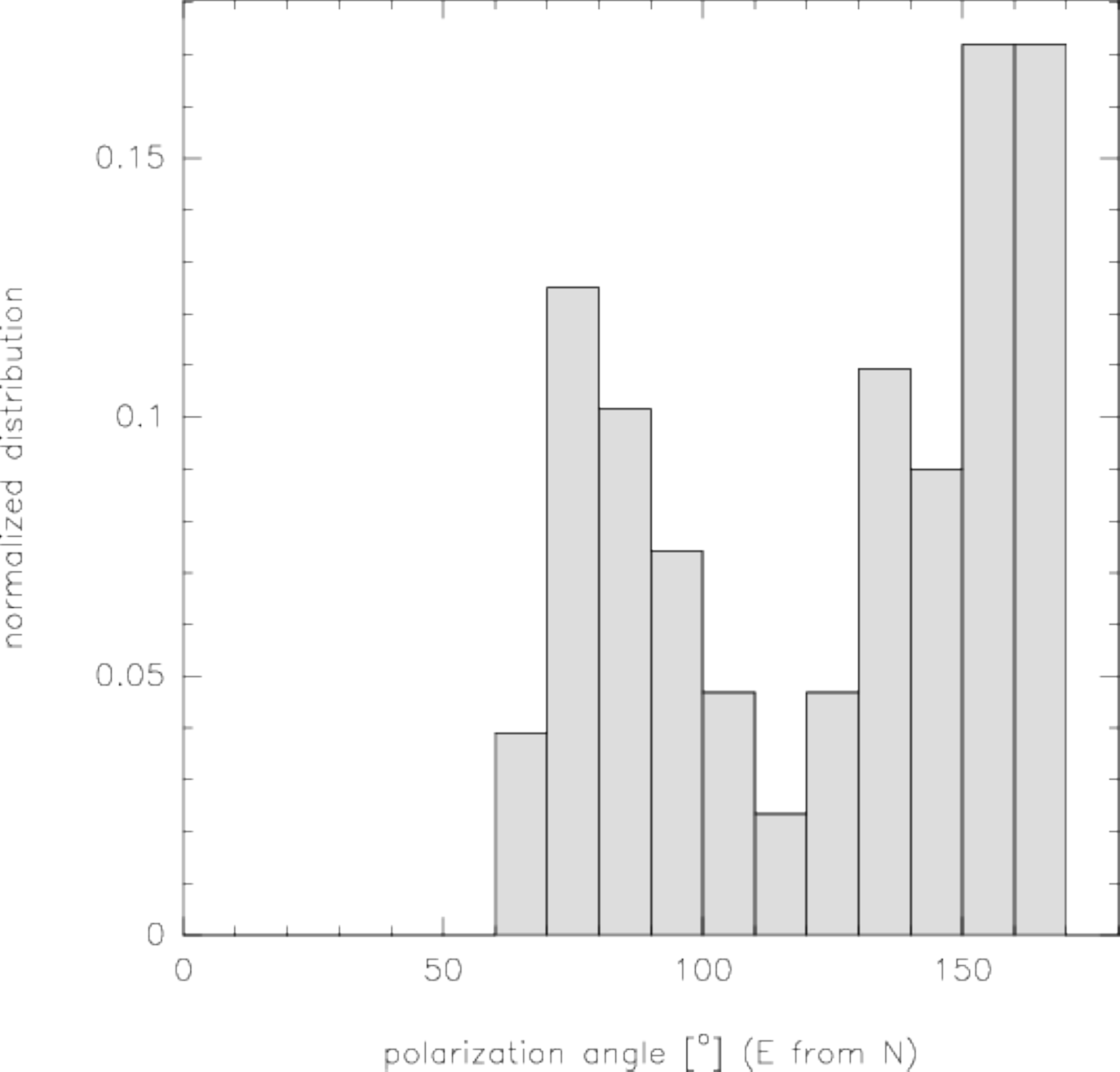} 
\includegraphics[scale=0.35,angle=0]{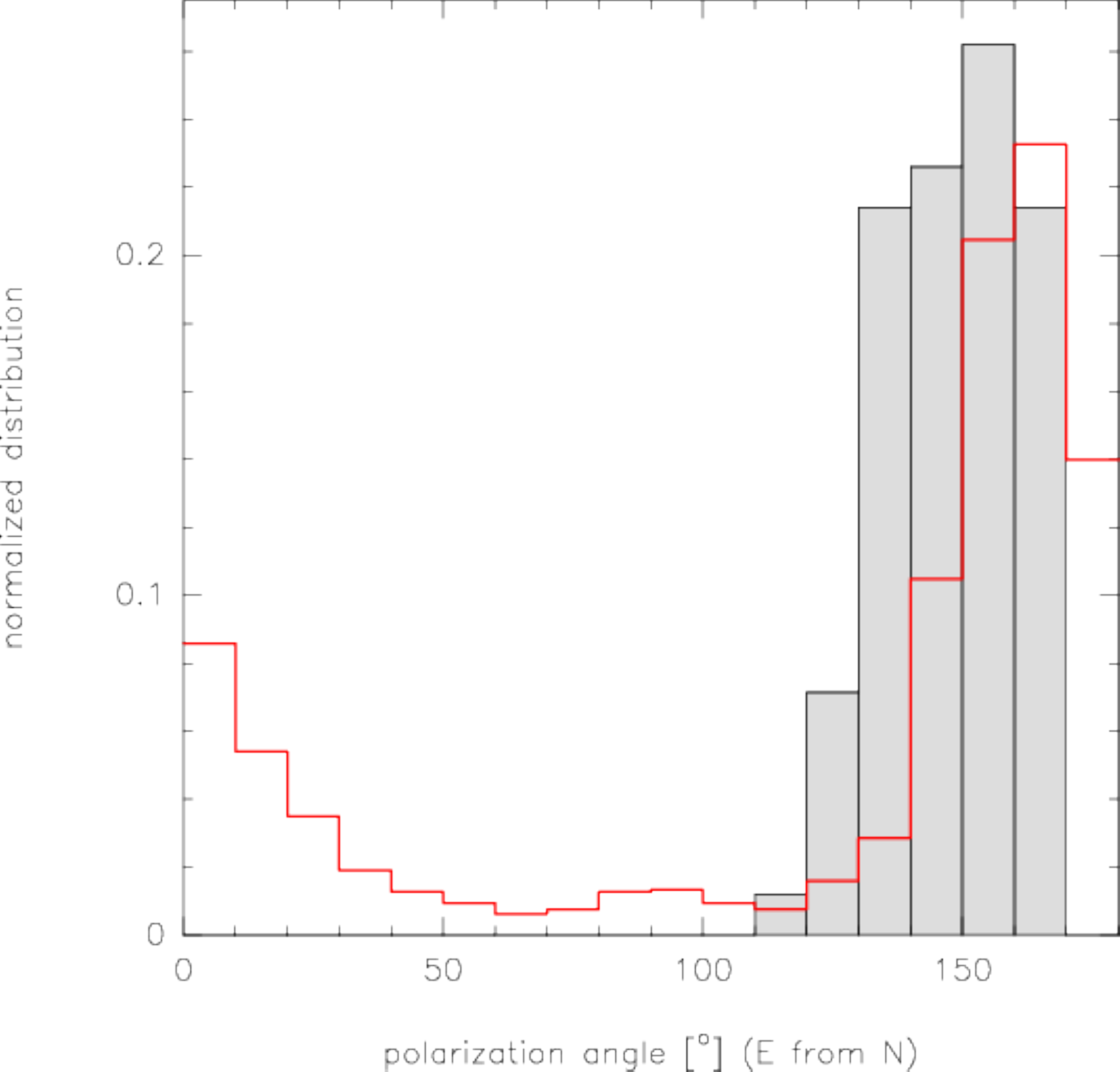} 
\caption{Distribution of polarization vectors in Tau~A.
Left: In the whole synchrotron nebula, above the 0.1~Jy/beam contour.
Right: in the inner $100\arcsec$ (filled gray histogram). For comparison, the corresponding
distribution for optical data from a central field of the same size
(HST/ACS, \citealp{2013MNRAS.433.2564M}) is also shown (red histogram).}
\label{fig:crabhisto}
\end{center}
\end{figure}

Our polarization map of Tau~A is well consistent with a 32~GHz polarization map from the Effelsberg
telescope, with a similar spatial resolution ($26\arcsec$ {\sc fwhm}, \citealp{1998MmSAI..69..933R}).
A further comparison with a 5~GHz VLA map \citep{2001ApJ...560..254B} with $1\farcs 4$ resolution
shows that the magnetic field orientation observed with PolKa is in the plane of the X-ray torus
\citep{2000ApJ...536L..81W}. South-east of
the torus, i.e. along the southern lobe of the jet, the polarization angles are similar and suggest here a
magnetic field that is toroidal with respect to the jet axis. As a matter of fact, the kinked jet was
successfully modeled by \citet{2013MNRAS.436.1102M} assuming such a magnetic field configuration, naturally
leading to the observed polarization signature. In the outer part the magnetic field structure is more complicated.
Remarkably, the distribution of polarization angles (Fig.~\ref{fig:crabhisto}) shows two peaks which correspond to
components that are roughly orthogonal. The peak at $160\dg$ corresponds to the emission near the torus, while the
other peak represents the body of the nebula. Our findings favor a scenario in
which the torus is magnetically confined because plasmas with crossed magnetic fields cannot penetrate each other 
\citep{2008ARA&A..46..127H}. Furthermore, a histogram (Fig.~\ref{fig:crabhisto}) of polarization angles in the
inner part of the synchrotron nebula, measured in the optical (HST/ACS, \citealp{2013MNRAS.433.2564M}), peaks at a
polarization angle of $150\dg$, which is within $10\dg$ from the peak in the corresponding distribution measured
with PolKa (the histogram of the optical data is wider because the sub-arcsecond resolution of the HST/ACS data
traces the polarization of individual filaments). We take these correspondences as genuine pieces of evidence that
PolKa reproduces the sky-plane magnetic field component of Tau~A correctly, and confirms that the observed
structures are magnetically controlled.
\begin{table*}[ht!]
\begin{center}
\caption{Synoptic summary of results.}
\label{tab:results}
\begin{tabular}{lccccccc}
\tableline\tableline
 & Stokes $I$  & \multicolumn{3}{c}{\Vhrulefill\,$\pL$ [\%]\,
\Vhrulefill} & \multicolumn{3}{c}{\Vhrulefill\,$\pa$ [$\dg$]\,\Vhrulefill} \\
 & [Jy] & {\sc polka} & {\sc xpol} & {\sc scupol} & {\sc polka} & {\sc xpol} & {\sc scupol} \\
\tableline
\multicolumn{8}{c}{Tau~A}  \\
\tableline
Pulsar &1.63 & 25.3\pmm 3.0 (20.9)& 14\pmm 1 & 14.3\pmm 1.8 & 145.1\pmm 3.3 (147.3)& 158.1\pmm 0.5 & 140.0\pmm 2.8 \\ 
Peak   &1.72 & 25.0\pmm 3.1 (23.9)& 25 & 18.7\pmm 1.5 & 151.7\pmm 3.5 (155.6)& 149.0\pmm 1.4 & 146.1\pmm 2.1\\ 
\tableline
\multicolumn{8}{c}{OMC1}  \\
\tableline
KL    & 103.8  & 0.7\pmm 0.2 (0.7) & & 0.7\pmm 0.1 & 32.8\pmm 7.6 (42.7) & & 40.8\pmm 5.4 \\ 
South &  58.3  & 0.7\pmm 0.1 (1.4) & & 4.7\pmm 0.2 & 27.5\pmm 5.1 (28.4) & & 25.9\pmm 0.9 \\ 
Bar E & 4.28   & 1.9\pmm 0.3 (2.0) & & & 143.1\pmm 4.1 (151.0) & & \\ 
Bar W & 3.78   & 2.8\pmm 0.5 (2.4) & & & 142.0\pmm 4.7 (149.9) & & \\ 
\tableline
\end{tabular}
\tablecomments{The PolKa data is corrected for instrumental polarization (uncorrected values are
given in brackets). Flux densities refer to a $20\arcsec$ beam ({\sc fwhm}), errors in Stokes $I$ are
dominated by systematics. The position of Orion KL is
$\alpha_{\rm J 2000} =$05:35:14.283, $\delta_{\rm J 2000} =$--05:22:31.32, in Tau~A the pulsar is at
$\alpha_{\rm J 2000} =$05:34:31.938, $\delta_{\rm J 2000} =$+22:00:52.18. Results from 30m/XPOL
(\citealp{2010A&A...514A..70A}, $27\arcsec$ {\sc fwhm}) and JCMT/SCUPOL (\citealp{2009ApJS..182..143M},
$20\arcsec$ {\sc fwhm}) are shown for comparison.}
\end{center}
\end{table*}
\subsection{OMC1 }
\label{subsec:omc1}
The first detection of the polarization of the dust emission from the Orion Molecular Cloud I
(OMC1) was made at \wl{270} with the Kuiper Airborne Observatory at a spatial resolution of
$90\arcsec$ (\citealp{1984ApJ...284L..51H}, further references to earlier attempts therein).
\citet{1998ApJ...493..811S} observed OMC1 at far infrared/submillimeter
wavelengths (\wl{100} and \wl{350}, respectively) across an $8\arcmin \times
8\arcmin$ large field. He confirmed the relatively weak linear polarization and 
its position angle measured by \citet{1984ApJ...284L..51H} towards the
Kleinman-Low nebula (Orion~KL) which is thought to be powered by an explosive
event \citep{2011A&A...529A..24Z}, while the neighboring high mass star-forming
region Orion-South and the envelope of Orion~KL exhibit a stronger polarization.
\citet{1998ApJ...493..811S} explains the depolarization in Orion~KL by the rising
dust opacity towards the far-infrared, while the low polarization in the Orion Bar,
a photon-dominated region seen edge-on \citep{1985ApJ...291..747T}, is attributed to 
a magnetic field pointing to the observer.
\citet{2008ApJ...679L..25V} studied the polarization spectrum of OMC-1 and conclude that a polarization
minimum occurs between \wl{100} and \wl{350} while \citet{2004ApJ...604..717H}, investigating the large
scale structure of the magnetic field in Orion~A, confirmed the relatively smooth polarization angle
structure towards OMC1 and interprete the weak polarization levels in the Orion bar with the lower
dust temperature in that region.

\begin{figure}
\includegraphics[scale=0.70,angle=0]{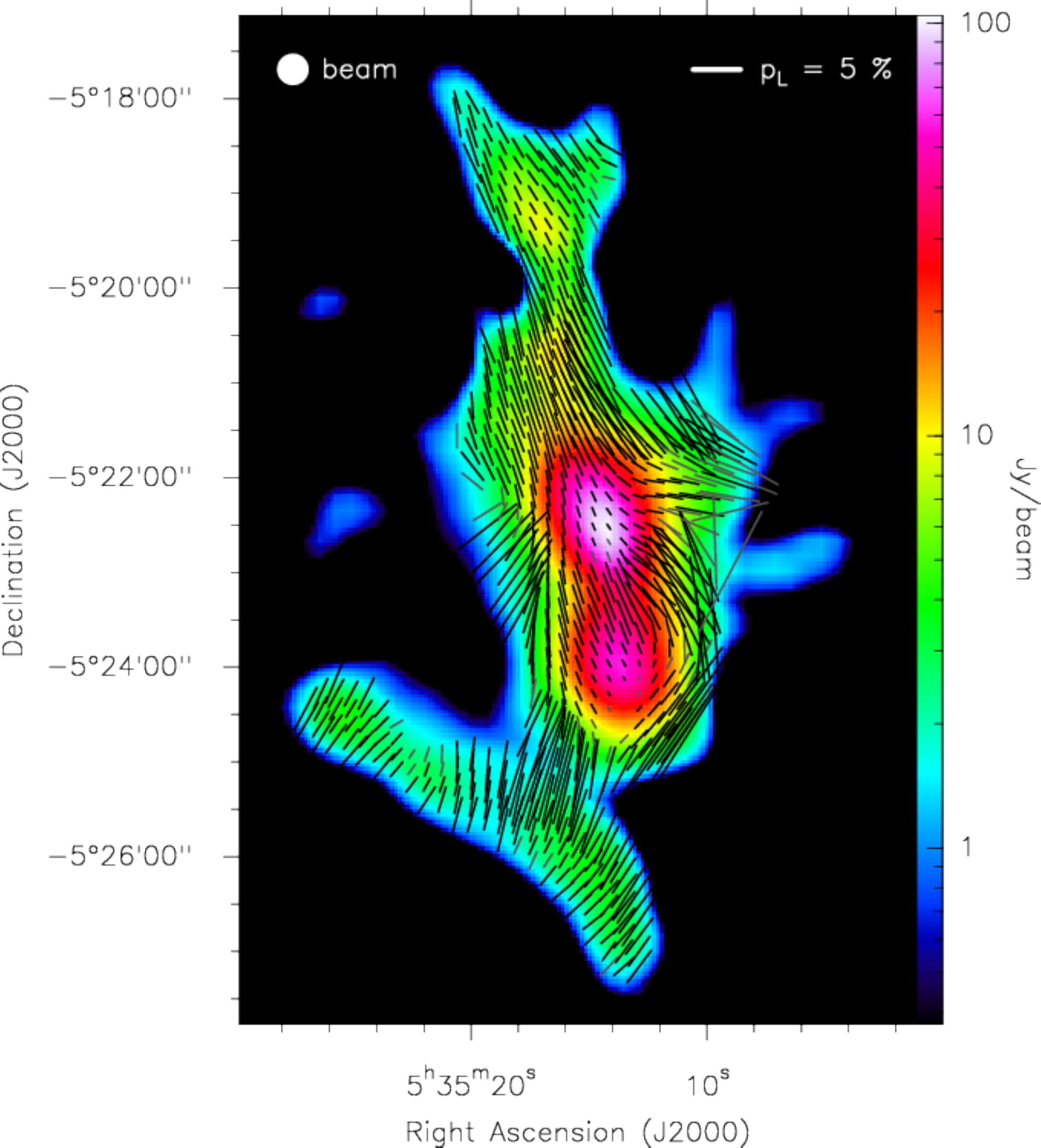}
\begin{center}
\caption{\wl{870} polarization vectors, corrected for instrumental
polarization, in OMC1 ($20\arcsec$~{\sc fwhm}, shown in the upper left corner)
with Stokes $I$ emission underneath (color plot scale to the right). A linear
polarization of 5\% is indicated in the top right corner. Use of black and gray
polarization vectors as in Fig.~\ref{fig:crab}.
}
\label{fig:omc1}
\end{center}
\end{figure}

Our polarization map of OMC1 is shown in Fig.~\ref{fig:omc1} and confirms the hourglass-like structure
of the magnetic field found by the aforementioned studies. The map 
is obtained from a total of 54 on-the-fly scans (i.e., a total of 2.25~h on source), 
and observed with a typical atmospheric transmission of 55 to 72\%. The sensitivity
across the map is $\sim30$mJy/beam. Our results agree with those from SCUPOL
\citep{2009ApJS..182..143M}, except for the fractional linear polarization that we
detect towards OMC1-South which is as weak as in the BN/KL region where it agrees
with SCUPOL ($\pL = 0.7$\%). The polarization angles in both regions are
$\sim$30$\dg$. We note that in general the correction for instrumental polarization
improves, across the map, the agreement between our results and those of SCUPOL.
The different levels of line contamination in SCUPOL and PolKa, contributing substantially to
Stokes $I$ but barely to $\pL$, may explain part of the differences. -- The polarization in
the filament extending northwards from Orion BN/KL is parallel to its long axis
(i.e., the projected magnetic field perpendicular to it), which is consistent with
previously reported results. We also note that our polarization angles agree,
within the errors, with those measured by \citet{1984ApJ...284L..51H} despite
the different wavelengths and spatial resolutions. We therefore confirm a fairly smooth magnetic field
structure where only at smaller spatial scales ($<20\arcsec$, i.e., 0.04~pc) depolarization occurs towards the cores
of Orion KL and South. This is presumably due to either a more complex magnetic field structure there or
a mixture of dust grains whose emission traces the same volume but have different properties such as temperature
and size distribution, as pointed out by \citet{2012ApJS..201...13V}. Their histogram of differential
polarization angles at $\lambda$350/850~$\mu$m indeed peaks in the $0\dg-10\dg$ interval.

As for the Orion Bar, a prototypical photon-dominated region, we find a fractional polarization that is significantly
higher, by a factor of 3 to 4, compared to the BN/KL and South peaks. It traces a magnetic field of which the sky-plane
projection is actually along the Bar, in disagreement with the suggestion by \citet{1998ApJ...493..811S}. It is
uncertain whether this result confirms the conjecture of \citet{2004ApJ...604..717H} that the dust temperature is
lower in the Bar than in BN/KL. From their CO excitation modeling, \citet{2012A&A...538A..12P} infer gas
temperatures in excess of 350~K and 250~K towards BN/KL and the Bar, respectively, corroborating the H$_2$ excitation
study of \citet{2009ApJ...701..677S}. An in-depth discussion of these findings is beyond the scope of the work at hand
and will be followed up in a forthcoming study.

\section{Conclusions and outlook}
\label{sec:conclusions}
In this work we demonstrated that PolKa, a reflection-type polarimeter installed in the warm optics of the 
bolometer camera LABOCA, can provide reliable measurements of the polarization of cosmic dust and synchrotron
emission at submillimeter wavelengths, across fields as large as $\sim$10$\arcmin$. Similar to a classical
waveplate polarimeter, the polarized fraction of the received power is detected by virtue of its
modulation when the plane of polarization is continuously rotated by the waveplate.
Standing waves in such a design are difficult to suppress, especially if space limitations in the
receiver cabin are an issue. However, we could show that a total power beating 
adding a deterministic, strictly harmonic signal to the modulation can be efficiently filtered out, preserving both the
amplitude and phase of the modulation, and therefore the information about the intrinsic polarization.
A prerequesite for this filter to work is to observe the source in the on-the-fly mode. This observing mode,
with either a linear or spiral stroke pattern, is the preferred method to obtain a critically sampled map with
the bolometer array whose pixels are separated by two full half-power widths. In order to demodulate the
signals and to obtain maps of the Stokes parameters $Q$ and $U$ we introduced a dedicated algorithm,
generalized synchronous demodulation. Moreover, a strategy to remove the spurious instrumental conversion
of Stokes $I$ into Stokes $Q$ and $U$ has been proposed and successfully applied to the observations.

Our results obtained for Tau~A and OMC1 agree reasonably well with previously published data,
while we uncover the polarization structure of the dust emission in the Orion Bar, a prototypical photon-dominated
region. This finding suggests that the projection of the magnetic field onto the plane of the sky is oriented along
the Bar. The physical implications of such a pseudo-2D configuration will be discussed in a forthcoming publication.
Towards the BN/KL region the polarization angles at \wl{870} and \wl{270} are comparable and rule out a
wavelength-dependent rotation of the linear polarization. The polarization structure of the \wl{870}
synchroton emission from the supernova remnant Tau~A confirms that its synchrotron nebula is magnetically controlled. 

Meanwhile PolKa has been used to observe the polarization of molecular clouds and the star-forming regions
they harbor, down to flux densities of a few 100~mJy (e.g., \citealp{2014arXiv1408.5133A}). These data
demonstrate that the high sensitivity of LABOCA allows for a $3\sigma$ detection of a fractional
linear polarization of 10\% in a $10\arcmin$ wide field after 2 hours on source, under decent (0.7~mm of water
vapor) yet frequent weather conditions. 

\acknowledgements{We owe the APEX observatory staff a debt of gratitude. We thank Dr. S. Risse
from the Fraunhofer IOF (Jena, Germany) for very fruitful discussions concerning the design of PolKa.
The analyzer grids were donated by Manfred Tonutti (RWTH Aachen University, Germany). H.W. acknowledges
helpful comments from C. Thum and insightful discussions with D. Muders, S. Heyminck, A. Lobanov and
M. Houde. The referee helped to improve the paper further. -- T. Hezareh's research was funded by the
Alexander von Humboldt foundation. The data reduction
software used the GILDAS and CFITSIO libraries (\verb+www.iram.fr/IRAMFR/GILDAS+ and
\verb+heasarc.gsfc.nasa.gov/docs/software/+). The opacity coefficients were obtained from the {\it am}
model (S. Paine, SMA technical memo \#152, 2012) via the {\it kalibrate} module of the KOSMA software, Department of Physics, Cologne University.}

\begin{appendix}
\section{Polarimetry in coherency matrix formulation}
\label{app:coherency}
The transfer of a radiation field of mixed polarization through a series of optical
devices can be conveniently described in the framework of the Jones calculus in 
a $\mathbb C_2$ vector space \citep{1941JOSA...31..488J} in combination with the coherency
matrix formulation used in modern optics (e.g., \citealp{1999prop.book.....B}).
The description of ensembles in quantum theory lends itself to the introduction of the Stokes
parameters as coefficients appearing in the expansion of the coherency matrix by
the identity matrix and the three Pauli spin matrices \citep{1954PhRv...93..121F}:
In quantum-electrodynamics there are two probabilities to consider, namely, the
probability for a photon to be in a given polarization state, and the probability
for this polarization state to be represented in an ensemble of photons. The density
matrix reads
\begin{equation}
\boldsymbol\rho = \frac{1}{2} \left( \begin{array}{ll}
                          I+Q  & U-iV \\
                          U+iV & I-Q
                          \end{array}
                   \right)\,.
\label{eq:rho}
\end{equation}
In wave optics, the Stokes parameters $I$, $Q$, $U$ and $V$ would be replaced by the corresponding integrals of Kirchhoff's
diffraction formula. Subsequent reflections and rotations of the plane of incidence are then decribed
by a series of similarity transformations ${\bf T} = {\bf T_1} \cdot {\bf T_2} \cdot {\bf T_3}$ etc. such that 
\begin{equation}
\boldsymbol\rho' = {\bf T} \boldsymbol\rho {\bf T^{\rm -1}}
\label{eq:similarityTransformation}
\end{equation} 
Like in quantum theory, the measurement is described by a projection operator,
i.e., an outer vector product
\begin{equation}
{\bf A} = 
\left(\begin{array}{c}
  g_{\rm x} \\
  g_{\rm y}
      \end{array}
\right)
  \cdot (g^*_{\rm x}, g^*_{\rm y}) =
\left(\begin{array}{ll}
      |g_{\rm x}|^2        & g_{\rm x}g^*_{\rm y} \\
      g^*_{\rm x} g_{\rm y}  & |g_{\rm y}|^2
      \end{array}
\right)
\label{eq:projOperator}
\end{equation}
where $g_{\rm x}$ and $g_{\rm y}$ are the complex gain factors for the signal detection in
horizontal, respectively vertical, polarization. The ensemble average $S$, i.e., the recorded signal,
can be shown to be
\begin{equation}
S = {\rm tr}\left(\boldsymbol\rho {\bf A} \right), 
\label{eq:measurementEquation}
\end{equation} 
which yields for, e.g., a receiver detecting horizontal polarization with
vanishing cross-polarization ($g_{\rm y} = 0$), $S = g_{\rm x}(I+Q)/2$,
as expected.

The coherency matrix formalism will now be applied to the measurement equation for a reflection-type polarimeter.
\citet{1996A&AS..117..137H} also combine the Jones and coherency matrix calculus but follow a mathematically different
approach. It can be shown that their description is formally equivalent to the treatment used here but the introduction 
of the projection operator may be more intuitive for the understanding of the underlying physical processes, since we
consider a detector as a filter for a given polarization state. 
\section{Derivation of the measurement equation}
\label{app:measurementEquation}
A half-waveplate rotates the plane of polarization by an angle $2\varphi$ without 
rotating the field of view. The corresponding transformation can therefore be
written as a similarity transformation of the coherency matrix as described
in the previous section. First, we rotate the coordinate system by an angle $\varphi$
such that the vertical axis of the new system is along the wires, i.e.,
\begin{equation}
{\bf T}_1 = \left(\begin{array}{cc}
              \cos{\varphi} & \sin{\varphi} \\
              -\sin{\varphi} & \cos{\varphi}
                    \end{array}
             \right)\,.
\label{eq:gridRotation}
\end{equation}
The next transformation describes the reflection off the wires of the
vertically polarized component (the first term on the r.h.s.), while the horizontally polarized component
is transmitted and reflected by the mirror, with a phase shift $\Phi$ (the second term):
\begin{equation}
{\bf T}_2 = \left(\begin{array}{cc}
             0 & 0 \\
             0 & 1 
                   \end{array}\
             \right) +
            \left(\begin{array}{cc}
             \exp{i\Phi}  & 0 \\
             0 & 0 
                   \end{array}\
             \right) . 
\label{eq:gridTransmission}
\end{equation}
Rotating back the coordinate system to the original orientation, and accounting
for the reflection of both polarization components is achieved by 
\begin{equation}
{\bf T}_3 = \left(\begin{array}{rr}
              -\cos{\varphi} & \sin{\varphi} \\
               \sin{\varphi} & \cos{\varphi}
                    \end{array}
             \right)\,.
\label{eq:gridRotaryReflection}
\end{equation}
Applying the transformation ${\bf T} = {\bf T}_3 {\bf T}_2 {\bf T}_1$ to the
density matrix equation~(\ref{eq:similarityTransformation}) yields (note that
the reflection of both polarization components changes the signs of Stokes $U$
and $V$)
\begin{equation}
\boldsymbol \rho' = \frac{1}{2}\left(\begin{array}{rr}
            I+Q' & -U'+iV' \\
         -U'-iV' & I-Q' 
              \end{array}
             \right)
\label{eq:rhoprim}
\end{equation}
with 
\begin{equation}
Q' = Q(\cos^2{2\varphi}+\sin^2{2\varphi}\cos{\Phi}), \ts
U' = \frac{1}{2}U(1-\cos{\Phi})\sin{4\varphi}, \ts
V' = \frac{1}{2}\sin{2\varphi}\sin{\Phi}\,.
\label{eq:rhoprimElements}
\end{equation}
If the wires of the analyzer grid are oriented vertically, then horizontal
polarization is transmitted and detected, i.e., the projection operator
equation~(\ref{eq:projOperator}) reads, for 100\% transmission,
\begin{equation}
{\bf A} = 
\left(\begin{array}{ll}
      1 & 0 \\
      0 & 0 
      \end{array}
\right)
\end{equation}
and equation~(\ref{eq:measurementEquation}) yields
$S=tr(\boldsymbol\rho' {\bf A}) =\boldsymbol\rho'_{11}$.

The beam is folded several times on its way from the focal plane to the camera.
The corresponding transformation can be included as a series of reflections,
each described by a transformation
\begin{equation}
{\bf T}_{\rm r} = \left(\begin{array}{rr}
             -\cos{\Theta} & -\sin{\Theta}  \\ 
             -\sin{\Theta} & \cos{\Theta}  
          \end{array}
             \right)
\end{equation}
where $\Theta$ is the position angle of the axis about which the mirror is
tilted (with respect to the $x$ axis in the Cassegrain coordinate system,
Fig.~\ref{fig:focalPlane}). In Eqs.~(\ref{eq:rhoprimElements}) this transformation
leads to the substitution $\varphi \rightarrow \varphi-\Theta/2$ and then, by
virtue of equation~(\ref{eq:measurementEquation}), to the result given by
equation~(\ref{eq:measurement}).
\section{Correction for instrumental polarization}
\label{app:ipcorr}
In the following algorithm we assume that the Stokes parameters are linear,
i.e., the instrumental polarization adds to the intrinsic
one. We neglect the spurious conversion among the Stokes parameters $Q$ and $U$,
which is caused by the error in the determination of the orientation of the
analyzer grid. In the following, $S_0, S_1, S_2$ and $S_3$ denote the brightness
distribution of the Stokes parameters $I$, $Q$, $U$ and $V$ on the sky. $P_1, P_2$
and $P_3$ describe the power pattern of the instrumental conversion from Stokes
$I$ into Stokes $Q$, $U$ and $V$, respectively, whereas $P_0$ is power beam pattern
of the antenna.

The ideal response of the telescope and its optics in the receiver cabin would
be
\begin{equation}
	F_j = S_j * P_0 \,\,\mbox{ for }j=0 \mbox{ to } 3
\end{equation}
such that the $F_{\rm j}$ are main-beam calibrated flux densities of the Stokes
parameters. In reality, the observations yield rather
\begin{equation}
	F_{\rm 0, obs} = F_0,\,\,F_{\rm j, obs} = S_{\rm j}*P_0 + S_0*P_{\rm j} \,\,\mbox{for }j=1, 2, 3
\end{equation}
All quantities in this set of equations are functions of offsets in the plane tangential
to the celestial sphere, i.e., ($\Delta\alpha \cos{\delta}, \Delta\delta$), and $*$ is the
convolution product. The basic idea of the correction procedure is that the
response functions $P_1, P_2$ and $P_3$ can be measured on a spatially
unresolved, unpolarized calibrator, e.g., Uranus, Mars or Mercury (the latter 
two are intrinsically weakly linearly polarized but the radial orientation
of polarization vectors leads to a mutual cancellation within the telescope's
main beam; Mercury should be observed at full phase). Such "Stokes beam maps",
sampled with the same mapping procedure as the maps to be corrected,
i.e., on-the-fly maps with a spiral or linear stroke pattern, yield
\begin{equation}
	F_{\rm j, cal} = \Pi * P_{\rm j}\,\,\mbox{for }j=1,2,3
\end{equation}
where the indices $j$ stand again for the Stokes parameters $Q$, $U$ and $V$.
The following analysis is computationally easier to perform in Fourier space
(no information will be lost, provided that aliasing in the discrete fast
Fourier transform is avoided by using sufficiently large maps or, if the
observed emission does not fall to zero within the map, to apply an apodization).
The 2D Fourier transforms, as a function of their spatial frequencies,
are denoted $\hat F_{\rm j}$, $\hat P_{\rm j}$ for $j=0,1,2,3$. From the Stokes 
$I$ image of our source and the "Stokes beams" we can model the instrumental
polarization
\begin{equation}
\hat F_{\rm j, mod} = \hat F_0 \cdot \hat F_{\rm j, cal}\,\,\mbox{for }
j=1,2,3
\end{equation}
and reconstruct $Q$ and $U$ from the Fourier transform of
\begin{equation}
\hat F_{\rm j}\cdot\hat\Pi\cdot\hat P_0
= \hat F_{\rm j, obs}\cdot \hat\Pi\cdot\hat P_0-\hat F_{\rm j, mod}
\end{equation}
We note that while the correction is done at a reduced spatial resolution, it
is possible to recover the original resolution by dividing by
$\hat Pi \cdot \hat P_0$ up to a reasonable cutoff of spatial frequencies,
and transforming the resulting $\hat F_{\rm j}$ back to the sky plane. This
means that we can only correct for the effects of instrumental polarization
down to spatial scales which are a factor $\sqrt{2}$ larger than the original
resolution. Only an interferometer map of $F_0$, with an order of magntitude
better resolution, can avoid this limitation (provided that the interferometer
data are corrected for the missing short spatial frequencies). 

Our algorithm has been demonstrated with the test source shown in
Fig.~\ref{fig:model} and also with XPOl data from the 30m telescope \citep{2013A&A...558A..45H}.
As expected, the correction is largest at the edges of
sources with a nearby strong emission peak, because the Stokes
beams $P_{\rm j}$ ($j=1,2,3$) are usually wider than the antenna power beam
pattern $P_0$, resulting in an increase of the fractional instrumental
polarization. This is often a direct consequence of the fact that the design of
telescopes and their tertiary optics is optimized for Stokes $I$ but not for the
other Stokes parameters.
\section{Low-pass wavelet filter}
\label{app:wavelet}
The properties of the discrete wavelet transformation (hereafter DWT, see e.g.,
\citealt{1992nrca.book.....P}) has properties that are similar to those of the
fast Fourier transform, e.g., the basis functions of both linear transformations
are localized in frequency space. The basis functions of the DWT, however, 
are also localized in the time domain.

\begin{figure}
\includegraphics[scale=0.3,angle=0]{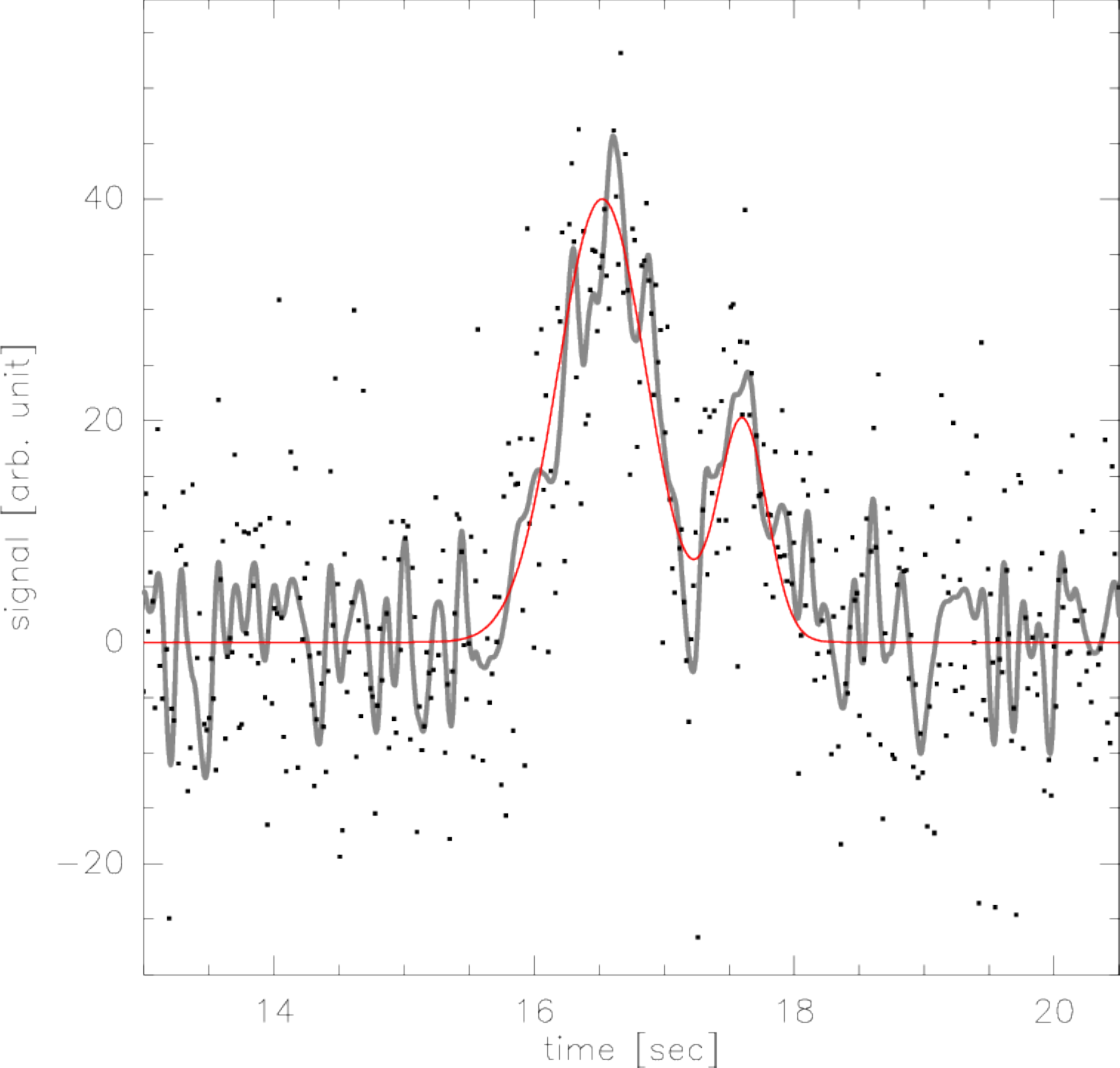}
\includegraphics[scale=0.3,angle=0]{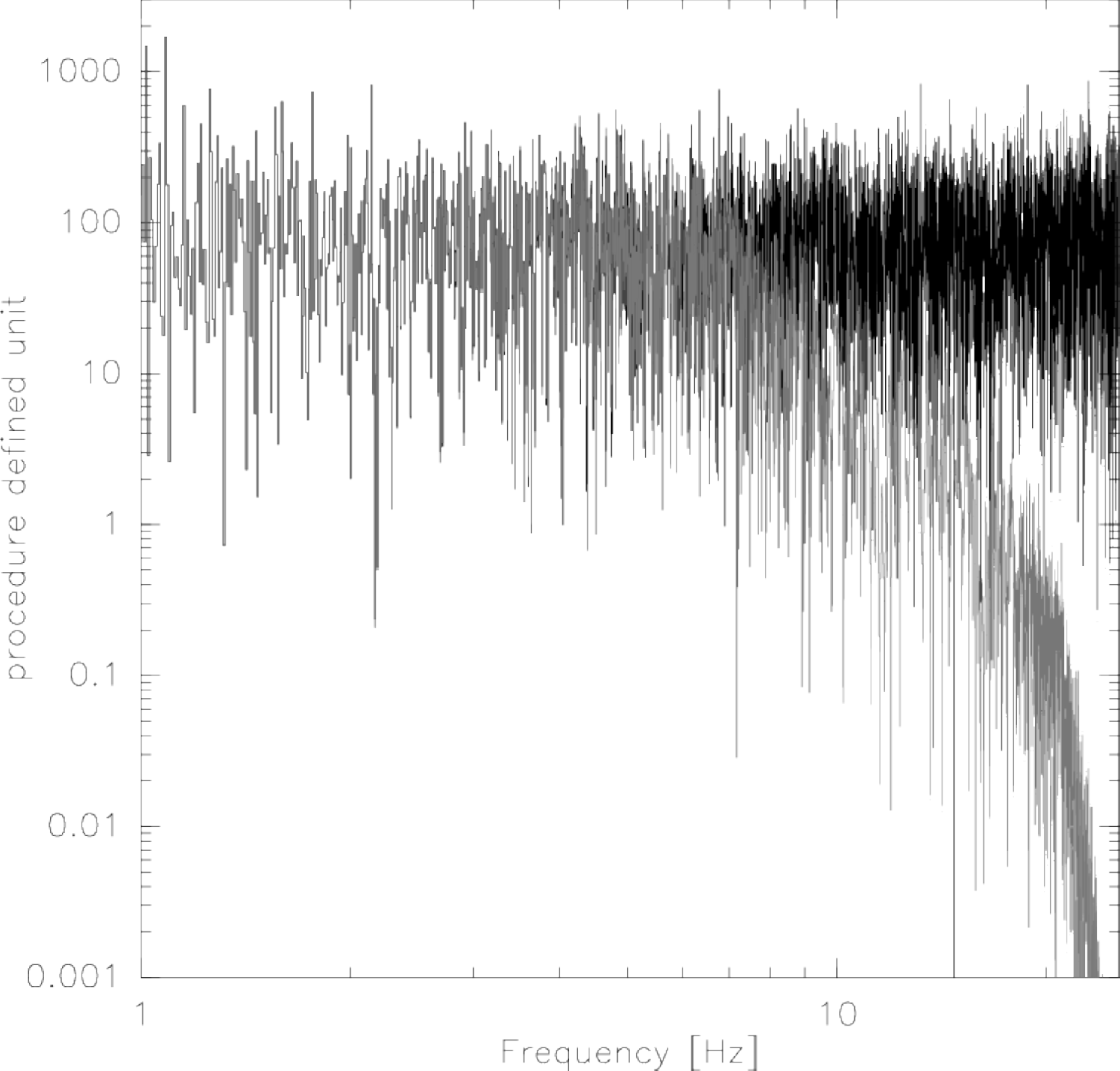}
\begin{center}
\caption{Removal of high-frequency noise with a wavelet filter. The test source has a double Gaussian profile and
is repeatedly scanned, like in a real on-the-fly map. The signal-to-noise ratio at peak is four. {\bf Left:} Part
of the modeled time series. The red line shows the input model before, the black dots after adding the Gaussian noise.
The thick gray line shows the profile after application of the wavelet filter. {\bf Right:} Spectral power density
before (black) and after (gray) application of the wavelet filter.}
\label{fig:dwt}
\end{center}
\end{figure}

Here we use the DWT to filter out the high-frequency noise in the time series
of signals measured by a bolometer pixel, applying Daubechies wavelet functions
with up to 12 coefficients \citep{1992tlw..conf.....D}. We perform the DWT,
identify the contribution of the high-frequency noise (contained in the high
wavelet numbers), set it to zero, and transform back to the time domain. 
Fig.~\ref{fig:dwt} shows a demonstration where only up to a quarter of the
total number ($2^{13}$) of wavelets is retained and transformed back. The
spectral analysis of the time series before and after the wavelet filtering
shows that the high frequency noise ($f > 8$~Hz) is efficiently suppressed,
while the intrinsic profile of the source is preserved.
\end{appendix}
\end{document}